\documentstyle[prb,aps,multicol,epsf]{revtex}

\begin{document}
\draft

\title{Theory of strong inelastic co-tunneling}

\author{A.~Furusaki\cite{AF} and K.~A.~Matveev}

\address{Department of Physics, Massachusetts Institute of Technology,
  Cambridge, MA 02139}

\date{August 3, 1995}
\maketitle

\begin{abstract}
  We develop a theory of the conductance of a quantum dot connected to
  two leads by single-mode quantum point contacts. If the contacts are in
  the regime of perfect transmission, the conductance shows no Coulomb
  blockade oscillations as a function of the gate voltage. In the presence
  of small reflection in {\it both\/} contacts, the conductance develops
  small Coulomb blockade oscillations. As the temperature of the system is
  lowered, the amplitude of the oscillations grows, and eventually sharp
  periodic peaks in conductance are formed. Away from the centers of the
  peaks the conductance vanishes at low temperatures as $T^2$, in
  agreement with the theory of inelastic co-tunneling developed for the
  weak-tunneling case. Conductance near the center of a peak can be
  studied using an analogy with the multichannel Kondo problem. In the
  case of symmetric barriers, the peak conductance at $T\to 0$ is of the
  order of $e^2/\hbar$. In the asymmetric case, the peak conductance
  vanishes linearly in temperature.
\end{abstract}

\begin{multicols}{2}

\section{Introduction}

Electronic transport in mesoscopic systems is usually affected by the
Coulomb interactions between the electrons. The interactions manifest
themselves most dramatically in small tunnel junctions. If a small
metallic grain is connected to a lead by a tunnel junction, an electron
tunneling into such a grain charges it by the elementary charge $e$ and
increases the energy of the system by $e^2/2C$. For a small grain, the
typical capacitance of the system can be very small, so that the
charging energy $e^2/2C$ can significantly exceed the temperature of the
system $T$. In this regime one observes the Coulomb blockade: strong
suppression of the tunneling into the grain at low temperatures. Various
aspects of the Coulomb blockade have been intensively studied in recent
years, both experimentally and theoretically.\cite{Averin,Grabert}

\begin{figure}
\narrowtext
\epsfxsize=3.1in
\hspace{.2em}
\epsffile{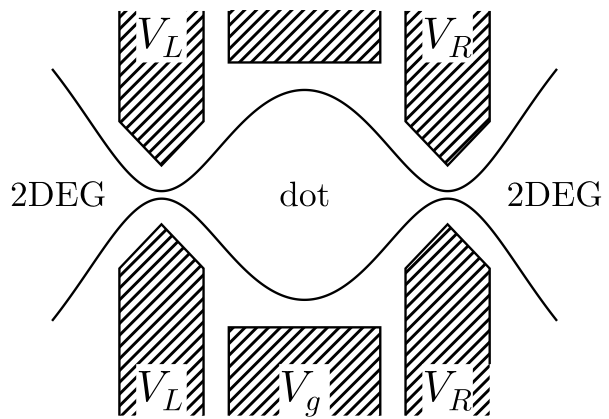}
\vspace{1.5\baselineskip}
\caption{Schematic view of a quantum dot connected to two bulk 2D
  electrodes. The dot is formed by applying negative voltage to the gates
  (shaded). Solid line shows the boundary of the 2D electron gas (2DEG).
  Electrostatic conditions in the dot are controlled by the voltage
  applied to the central gates. Voltage $V_{L,R}$ applied to the auxiliary
  gates controls the transmission probability through the left and right
  constrictions.}
\label{fig:1}
\end{figure}

Recently it has become possible\cite{Kastner} to observe the Coulomb
blockade in semiconductor heterostructures. Instead of a metallic grain,
in such experiments one creates a small region of the two-dimensional
electron gas (2DEG)---a quantum dot---separated from the rest of the
2DEG by potential barriers created by applying negative voltage to the
gates, Fig.~\ref{fig:1}.  Experimentally, one measures the linear
conductance through the quantum dot as a function of the voltage $V_g$
applied to the central gates. The role of the gate voltage is to change
the electrostatic energy of the system,
\begin{equation}
E = \frac{(Q-eN)^2}{2C},
  \label{electrostatic_energy}
\end{equation}
where $Q$ is the charge of the quantum dot, and $N$ is a dimensionless
parameter proportional to the gate voltage and the capacitance $C_g$
between the dot and the gate, $N=C_g V_g/e$. If the potential barriers
separating the quantum dot from the leads are high, the charge $Q$ is
integer in units of $e$. Thus, the ground state of the system corresponds
to some integer value of charge, and the states with different values of
$Q$ are separated from it by the electrostatic gap $\Delta\sim e^2/2C$. It
is important to note that at the values of the gate voltage corresponding
to half-integer values of $N$ the gap vanishes, i.e., the ground state is
degenerate in charge $Q$. For example, at $N=\frac{1}{2}$ the ground state
can have either charge $Q=0$ or $e$. Therefore at half-integer $N$ the
tunneling of an electron into or out of the quantum dot does not lead to
the increase of the electrostatic energy of the system, and the Coulomb
blockade is lifted. As a result the conductance of the structure in
Fig.~\ref{fig:1} shows periodic peaks as a function of $N$. The Coulomb
blockade peaks in linear conductance as a function of the gate voltage are
readily observed in the experiments (see, e.g.,
Ref.~\onlinecite{Kastner}).

\vfill
\vspace{\baselineskip}

The shapes and heights of the peaks in conductance are usually described
within the framework of the orthodox theory of the Coulomb
blockade.\cite{Averin} In this approach the conductance is found from the
balance of the rates of tunneling of electrons between the leads and the
dot, with the rates being calculated in the lowest order perturbation
theory in the tunneling matrix elements. The resulting linear conductance
has the form:\cite{Glazman}
\begin{equation}
  G = \frac{1}{2}\frac{G_LG_R}{G_L + G_R}
      \frac{\Delta/T}{\sinh(\Delta/T)},
  \quad \Delta=\frac{e^2}{C}\delta N.
  \label{orthodox}
\end{equation}
Here $G_L$ and $G_R$ are the conductances of the left and right barriers,
$\delta N$ is the deviation of $N$ from the nearest half-integer value.
Away from the center of the peak, the tunneling of an electron into the
quantum dot is suppressed, because only a small fraction of electrons has
the energy sufficient to overcome the electrostatic gap $\Delta$. Thus the
conductance (\ref{orthodox}) decays exponentially at $T\to0$. At very low
temperatures, however, another mechanism---usually referred to as {\it
  inelastic co-tunneling}---dominates the conductance.  Instead of a
real process of tunneling from a lead to the dot, electron tunnels from
the left lead to a virtual state in the dot, and then another electron
tunnels from the dot to the right lead. As a result of such a two-step
process, one electron is transferred from the left lead to the right one,
and the charge of the quantum dot is unchanged. The corresponding
conductance has only a power-law dependence on temperature,\cite{AN}
\begin{equation}
G=\frac{\pi\hbar G_LG_R}{3e^2}\left(\frac{T}{\Delta}\right)^2,
\quad
T\ll\Delta.
\label{cotunneling}
\end{equation}
In the case of weak tunneling, $G_L, G_R\ll e^2/\hbar$, the results
(\ref{orthodox}) and (\ref{cotunneling}) give the full description of the
Coulomb blockade peaks in linear conductance. Near the center of the peak
the conductance is given by Eq.~(\ref{orthodox}), whereas the tails of the
peaks are dominated by the co-tunneling contribution (\ref{cotunneling}).

In a number of recent experiments\cite{van,Pasquier,Waugh,Molenkamp} the
tunneling through quantum dots was studied in the regime when one or more
quantum point contacts were tuned to the regime of strong tunneling,
corresponding to conductance $G\to e^2/\pi\hbar$. In particular, it was
demonstrated by van der Vaart {\it et al.\/}\cite{van} that the Coulomb
blockade oscillations of linear conductance persist as long as the
conductances of both barriers are below the conductance quantum
$e^2/\pi\hbar$.  Most of the existing theoretical works on the
strong-tunneling regime of the Coulomb
blockade\cite{GS,PZ,Falci,Schoeller} discuss the case of metallic
systems, where one can achieve large conductance of a tunnel junction
between the grain and a lead by increasing the number of transverse modes
while keeping the transmission coefficient ${\cal T}$ small. Such theories
cannot be applied to the experiments in semiconductor heterostructures,
where the point contacts typically allow only a single transverse mode,
but the transmission coefficient can be tuned to ${\cal T} \to 1$. The
theoretical works for this case either do not concentrate on the Coulomb
blockade oscillations\cite{Flensberg} or study the oscillations of
equilibrium properties of the system,\cite{Matveev1} rather than its
conductance.

In this paper we develop a theory of the Coulomb blockade oscillations of
conductance in the regime of strong tunneling. It was shown in
Ref.~\onlinecite{Matveev1} that the average charge of a quantum dot
connected to a lead by a single-mode quantum point contact shows small
periodic oscillations as a function of the gate voltage as long as the
transmission coefficient ${\cal T}$ is below unity. Similarly, we will
show that the conductance through a quantum dot shows small periodic
oscillations at ${\cal T} < 1$. One should note, however, that unlike the
average charge,\cite{Matveev1} the amplitude of the oscillations of conductance
depends
on temperature and grows as $T$ is lowered. In the limit $T\to 0$, small
oscillations develop into sharp periodic peaks, and the off-peak
conductance vanishes as $T^2$, thus reproducing the characteristic
quadratic temperature dependence for inelastic co-tunneling
(\ref{cotunneling}).

In section \ref{weak} we use the analogy\cite{GM,Matveev2} between the
Coulomb blockade and multichannel Kondo problem to discuss the peak value
of conductance at $T\to0$. This discussion is done for the weak-tunneling
case, but the conclusion that the peak conductance is of the order of
$e^2/\hbar$ is valid for the strong tunneling as well. The theory of
conductance oscillations in the strong-tunneling regime is developed in
sections \ref{spinless} and \ref{withspin}. In section \ref{scaling} we
discuss the scaling properties of our problem. In particular, we show that
the quadratic temperature dependence of the off-peak conductance at
$T\to0$ is universal and holds for arbitrary transmission coefficients.

\section{Conductance peaks in the weak-tunneling case}
\label{weak}

Let us first consider the weak-tunneling case. The linear conductance is
given by Eqs.~(\ref{orthodox}) and (\ref{cotunneling}). One should note
that in the derivation of Eqs.~(\ref{orthodox}) and (\ref{cotunneling})
only the lowest-order terms of the perturbation theory in tunneling
amplitudes were taken into account. This is usually justified at $G_L, G_R
\ll e^2/\hbar$. However, it was shown in Ref.~\onlinecite{GM} that near a
half-integer $N$ and at low temperatures the higher-order corrections lead
to large logarithmic renormalizations of the tunneling amplitudes. The
nature of these renormalizations can be understood in terms of the
analogy\cite{Matveev2} between the Coulomb blockade and the Kondo problem.
We will now discuss this analogy and correct the results (\ref{orthodox})
and (\ref{cotunneling}) to take into account the renormalizations of the
tunneling amplitudes.

For simplicity we start with the quantum dot weakly coupled to a {\it
  single\/} lead. Such a system can be described by the following
Hamiltonian:
\begin{eqnarray}
H &=& \sum_{k} \epsilon_k a_k^\dagger a_k^{}
      + \sum_{p} \epsilon_p a_p^\dagger a_p^{}
      + E_C (\hat n - N)^2
\nonumber\\
  & & +\sum_{kp} (v_t^{} a_k^\dagger a_p^{} + v_t^* a_p^\dagger a_k^{}).
  \label{tunnel_hamiltonian}
\end{eqnarray}
Here $a_k$ and $a_p$ are the annihilation operators for the electrons in
the lead and in the dot respectively, and $E_C$ is the characteristic
charging energy, $E_C = e^2/2C$; coupling between the dot and the lead is
described by tunneling Hamiltonian with the matrix element $v_t^{}$.
Finally, $\hat n$ is the operator of the number of electrons in the dot,
counted from that at the ground state of the system without tunneling,
$\hat n = \sum (a_p^\dagger a_p^{} - \langle a_p^\dagger a_p^{}
\rangle_0^{})$.

Due to the last term in Eq.~(\ref{tunnel_hamiltonian}) electrons can
tunnel from the lead to a virtual state in the dot. The energy of such a
virtual state is of the order of $E_C$, and to complete this scattering
process an electron has to tunnel from the dot to the lead. It is
instructive to calculate the second-order amplitude of scattering between
two states in the lead, $k$ and $k'$, near the Fermi level:
\begin{eqnarray}
  T_{k\to k'} = \sum_p && \left[
                  \frac{|v_t^{}|^2 \theta(\epsilon_p)}
                       {E_C N^2 - E_C(1-N)^2 - \epsilon_p}
\right.\nonumber\\
              && \left.
                  - \frac{|v_t^{}|^2 \theta(-\epsilon_p)}
                       {E_C N^2 - E_C(1+N)^2 + \epsilon_p}
                       \right].
  \label{2_order_sum}
\end{eqnarray}
Here the second term corresponds to the process in which first an electron
from the filled state $p$ in the dot tunnels to the state $k'$ in the
lead, and then the electron $k$ fills the hole in the dot. Clearly, the
total amplitude diverges logarithmically at $N\to\pm\frac12$,
\begin{equation}
  T_{k\to k'} = -\nu_d |v_t^{}|^2 \ln\frac{1+2N}{1-2N},
  \label{2_order_log}
\end{equation}
where $\nu_d$ is the density of states in the dot. The logarithmic
singularity at $N\to\frac12$ originates from the degeneracy of the
electrostatic energy for states with $n=0$ and $n=1$ extra electrons in
the dot.

Let us now concentrate on the case of the dimensionless gate voltage $N$
very close to $\frac{1}{2}$. In this case the logarithmic renormalizations
are large, and to find the scaling properties of the scattering
amplitudes, we can restrict our consideration to the energy scales smaller
than $E_C$. Thus when constructing the perturbation series for a tunneling
amplitude we will neglect all the processes in which the system visits
virtual states with the charge $n$ different from 0 or 1.  Another
important property of the Hamiltonian (\ref{tunnel_hamiltonian}) is that
the perturbation---the tunneling term---always changes the charge of the
dot by 1. Suppose that at the first step of a high-order tunneling process
an electron tunnels from the lead to the dot thus changing $n$ from 0 to
1. Then at the next step an electron will have to tunnel from the dot to
the lead (changing $n=1\to0$), because otherwise a virtual state with
$n=2$ and correspondingly large energy $\sim E_C$ would have been created.
At the third step an electron tunnels from the lead to the dot, changing
$n=0\to1$, and so on.

One can easily see that exactly the same perturbation series for
scattering amplitudes is obtained from the Hamiltonian of the anisotropic
Kondo model,
\begin{eqnarray}
H &=& \sum_{k\alpha} \epsilon_k a_{k\alpha}^\dagger a_{k\alpha}^{}
      + h S^z
\nonumber\\
  & & +J_\perp\! \sum_{kk'\alpha\alpha'}\!
     (\sigma^{+}_{\alpha\alpha'} S^{-} + \sigma^{-}_{\alpha\alpha'} S^{+})
     \,a_{k\alpha}^\dagger a_{k'\alpha'}^{},
  \label{Kondo}
\end{eqnarray}
where $\sigma^{\pm}$ are the linear combinations of Pauli matrices,
$\sigma^{\pm}= \sigma^x \pm \sigma^y$, and {\boldmath $S$} is the impurity
spin, $S^{\pm}=S^x\pm S^y$.  A formal discussion of this mapping can be
found in Ref.~\onlinecite{Matveev2}; here we only mention the relations
between the parameters of the Hamiltonians (\ref{tunnel_hamiltonian}) and
(\ref{Kondo}). The values of the electron spin $\alpha=\ \uparrow$ and
$\downarrow$ correspond to the electron being in the lead and the dot
respectively. Similarly, the values of the impurity spin $S^z=\
\downarrow$ and $\uparrow$ correspond to the states with the number of
particles in the dot $n=0$ and 1. Magnetic field $h$ is identified with
the energy splitting of the $n=0$ and $n=1$ states, $h=2E_C(\frac12-N)$.
Finally, the tunneling matrix element $v_t^{}$ is mapped to the exchange
constant $J_\perp$.

It is well known that at low temperatures and magnetic fields the exchange
constant $J$ of the Kondo model renormalizes logarithmically. As a result
$J$ grows, and the system approaches the strong-coupling fixed point.  In
the context of the Coulomb blockade problem, this means that at
$N\to\frac12$ and $T\to0$ the tunneling amplitudes grow until the
conductance of the barrier becomes of order $e^2/\hbar$.

Using the analogy with the Kondo problem, one can actually find the
renormalizations of the tunneling amplitudes. In the leading logarithm
approximation one can use the poor man's scaling technique.\cite{Anderson}
The result\cite{GM} can be written down in terms of the renormalization of
the transmission coefficient ${\cal T}$ of the barrier between the lead
and the dot,
\begin{equation}
{\cal T} = \frac{{\cal T}_0}
           {\cos^2\bigl(\frac{1}{\pi}\sqrt{{\cal T}_0}\xi\bigr)},
\quad
\xi = \ln\frac{E_C}{\max\{T,\Delta\}}.
  \label{poor_man}
\end{equation}
Here ${\cal T}_0$ is the bare value of the transmission coefficient, and
$\Delta = E_C|1-2N|$.  When ${\cal T}$ becomes of the order of unity, the
leading logarithm approximation fails. Thus the result (\ref{poor_man}) is
valid as long as the renormalized transmission coefficient is small,
${\cal T}\ll 1$.

In this regime one can still use the weak-tunneling results
(\ref{orthodox}) and (\ref{cotunneling}) for the conductance through a dot
connected to two leads, taking into account the renormalizations of the
conductances of the barriers. This can be done by noting the relation
between the conductance of a barrier $G_b$ and its transmission
coefficient,
\begin{equation}
G_b=\frac{e^2}{2\pi\hbar}\times
    \begin{cases}{
        {\cal T}, & without spins,\cr
        2{\cal T}, & with spins.
     }\end{cases}
  \label{barrier_conductance}
\end{equation}
It is worth mentioning that within the leading logarithm approximation the
transmission coefficients of two barriers renormalize independently, so
one can use Eq.~(\ref{poor_man}) for both left and right barriers.

Let us find the temperature dependence of the peak value of the
conductance in the case of symmetric barriers, i.e., assume $N=\frac12$
and $G_L = G_R = G_b = (e^2/\pi\hbar) {\cal T}$. According to
Eq.~(\ref{orthodox}), the peak conductance is $\frac14 G_b$. Taking into
account the renormalization (\ref{poor_man}) of the transmission
coefficient, we find
\begin{equation}
G = \frac{e^2}{4\pi\hbar}
    \frac{{\cal T}_0}
           {\cos^2\left[\frac{1}{\pi}\sqrt{{\cal T}_0}\ln(E_C/T)\right]}.
  \label{peak_conductance}
\end{equation}
As expected, when the temperature is lowered, the conductance grows.  This
result is valid only at temperatures exceeding the characteristic Kondo
temperature of this problem, $T_K\simeq E_C\exp(-\pi^2/2\sqrt{{\cal
    T}_0})$. At too low temperatures, $T\lesssim T_K$, the conductance is
of the order of $e^2/\hbar$, transmission coefficient ${\cal T} \sim 1$,
and the result of the leading logarithm approximation
(\ref{peak_conductance}) is no longer applicable.

It is also interesting to study the tails of the conductance peaks. If the
gate voltage $N$ is not precisely half-integer, then at low temperatures,
$T\ll\Delta=2E_C|N-\frac12|$, one can still use the inelastic co-tunneling
result (\ref{cotunneling}) with the values of conductances of the barriers
renormalized according to Eqs.~(\ref{poor_man}) and
(\ref{barrier_conductance}). The resulting conductance has the form:
\begin{equation}
G=\frac{e^2{\cal T}_L{\cal T}_R}{3\pi\hbar}
  \frac{T^2}
   {\Delta^2
    \cos^2\left[\frac{1}{\pi}\sqrt{{\cal T}_L}\ln\frac{E_C}{\Delta}\right]
    \cos^2\left[\frac{1}{\pi}\sqrt{{\cal T}_R}\ln\frac{E_C}{\Delta}\right]}.
  \label{tails}
\end{equation}
Similarly to Eq.~(\ref{peak_conductance}), the result (\ref{tails}) is
valid only as long as ${\cal T}\ll 1$. In this case the condition allows
only the values of the gate voltage not too close to the resonance,
$\Delta\gg T_K$.

As we have seen, if one starts with the weak-tunneling case, in the
vicinity of the resonances in $G(V_g)$ the tunneling amplitude grows, and
the system approaches the strong-coupling regime. Therefore one can use
the results of the weak-tunneling approach to find the universal features
of the peaks in conductance which should also persist in the
strong-tunneling case. Such features are: (i) the $T\to0$ value of the
peak conductance (\ref{peak_conductance}) for symmetric barriers is of the
order of $e^2/\hbar$, and (ii) away from the centers of peaks the
temperature dependence of the conductance at $T\to0$ is $G\propto T^2$. It
is worth noting that unlike the temperature dependence of the co-tunneling
conductance (\ref{cotunneling}), its $1/\Delta^2$ dependence of the gate
voltage is not universal, see Eq.~(\ref{tails}).

\section{Strong co-tunneling of spinless electrons}
\label{spinless}

We will start our treatment of the strong-tunneling case with a simplified
model of spinless electrons. As we will see, in this case a complete
solution of the problem is possible. Experimentally, the spinless case can
be achieved by applying an in-plane magnetic field, sufficiently strong to
polarize the electron gas in the constrictions.

\subsection{Theoretical model}

We assume that each of the quantum point contacts in Fig.~\ref{fig:1}
allows the transmission of electrons in only one mode. The transmission
coefficients of the two contacts ${\cal T}_L$ and ${\cal T}_R$ are assumed
to be close to unity. This enables us to start with the Hamiltonian
corresponding to perfect transmission, ${\cal T}_L = {\cal T}_R = 1$, and
then construct the perturbation theory in small reflection amplitudes,
$r_L$ and $r_R$. Following Ref.~\onlinecite{Flensberg}, we will describe
each of the quantum point contacts by a system of one-dimensional (1D)
electrons. The formal derivation of this model\cite{Matveev1} is based on
the fact that if the boundaries of the channel are very smooth, the
transport though it can be described in terms of adiabatic
wavefunctions\cite{Lesovik} having essentially a 1D form.

Another simplification of the problem is due to the fact that the typical
energies involved in the Coulomb blockade phenomenon are of the order of
the charging energy $E_C$, which is much smaller than the Fermi energy.
This allows us to linearize the spectra of the 1D electrons near the
Fermi points and to formulate the model in terms of the left- and
right-moving fermions:
\begin{eqnarray}
H_0\! =\! i\hbar v_F\!\int\!\! &&
        \left[
         \psi^\dagger_{L1}(x)\nabla\psi_{L1}(x)
         \!-\!\psi^\dagger_{L2}(x)\nabla\psi_{L2}(x)
        \right.
\nonumber\\
      &&\left.\!\!\!
         +\psi^\dagger_{R1}(x)\nabla\psi_{R1}(x)
         \!-\!\psi^\dagger_{R2}(x)\nabla\psi_{R2}(x)
        \right] \!dx.
  \label{H_0_fermionic}
\end{eqnarray}
Here $\psi_{L1}$ and $\psi_{L2}$ are the annihilation operators for the
electrons of the left contact moving to the left and right respectively;
$\psi_{R1}$ and $\psi_{R2}$ describe the left- and right-moving electrons
of the right contact; $\nabla\equiv\frac{\partial}{\partial x}$. We
associate the centers of the constrictions with points $x=0$, meaning that
electrons in the dot are described by $\psi_{L1(2)}$ at $x>0$ and by
$\psi_{R1(2)}$ at $x<0$. Thus the charging energy can be presented as
\begin{equation}
  H_C = E_C(\hat n - N)^2,
\label{H_C}
\end{equation}
where
\[
\hat n = \int_{0}^{\infty}\!\sum_{j=1,2}
         \!\left[:\!\psi^\dagger_{Lj}(x)\psi_{Lj}^{}(x)\!\!:
         +:\!\psi^\dagger_{Rj}(-x)\psi_{Rj}^{}(-x)\!\!:\right]\!dx.
\]
Here $:\!\hat X\!\!:$ denotes the normal-ordered operator $\hat X$.
Finally, the presence of weak backscattering in the contacts, described by
small reflection amplitudes $|r_L|, |r_R| \ll 1$, gives rise to the term
mixing the left- and right-moving particles,
\begin{equation}
H' = \hbar v_F\!\left[
    |r_L|\psi^\dagger_{L1}(0)\psi_{L2}^{}(0)
   +|r_R|\psi^\dagger_{R1}(0)\psi_{R2}^{}(0)
   +\mbox{h.c.}\right]\!.
  \label{backscattering_fermionic}
\end{equation}

The Hamiltonian $H=H_0+H_C+H'$ gives the complete description of the
transport through a quantum dot connected to two leads. In our model the
two contacts are described by two independent 1D systems. Thus we neglect
the possibility of coherent transport of electrons from one quantum point
contact to the other. Such processes lead to the additional mechanism of
transport through the dot called {\it elastic co-tunneling.\/} It is
known,\cite{AN} however, that at temperatures exceeding the level spacing
$\varepsilon$ in the dot the elastic co-tunneling contribution to the
total conductance is much smaller than that of inelastic co-tunneling.
Thus we will concentrate on the case $T\gg\varepsilon$ which is easy to
satisfy in the experiments, and in the following calculations we will
assume $\varepsilon=0$. At $T\lesssim\varepsilon$ the conductance becomes
temperature-independent and can be estimated by replacing $T$ by
$\varepsilon$ in our results.

The solution of the problem with the Hamiltonian given by
Eqs.~(\ref{H_0_fermionic})--(\ref{backscattering_fermionic}) is not
trivial because the interaction term $H_C$ is not quadratic in fermion
operators. To overcome this difficulty, we will use the bosonization
technique\cite{Sol} which enables one to treat the interaction in 1D
systems exactly. Following Ref.~\onlinecite{Sol}, we present the fermion
operators as
\begin{equation}
  \begin{array}{rcl}
     \psi_{L1}(x) &=& \left(\frac{D}{2\pi\hbar v_F}\right)^{1/2}
                 \eta_{L1}^{} e^{i\varphi_{L1}^{}(x)},\\
     \psi_{L2}(x) &=& \left(\frac{D}{2\pi\hbar v_F}\right)^{1/2}
                 \eta_{L2}^{} e^{i\varphi_{L2}^{}(x)}.
  \end{array}
  \label{bosonization}
\end{equation}
Here $D$ is the bandwidth, and the two independent bosonic fields
$\varphi_{L1}^{}$ and $\varphi_{L2}^{}$ satisfy the following commutation
relations:
\begin{equation}
[\varphi_{L1}^{}(x), \varphi_{L1}^{}(y)]
 = -[\varphi_{L2}^{}(x), \varphi_{L2}^{}(y)]
 = -i\pi{\rm sgn}(x-y).
  \label{commutators}
\end{equation}
Given (\ref{commutators}), the operators (\ref{bosonization}) satisfy
usual anticommutation relations. To ensure proper anticommutation
relations for fermions at different branches, we have added
the local Majorana fermions $\eta_{L1}^{}$ and $\eta_{L2}^{}$.

The electrons of the right contact are
bosonized in a similar way. One can then rewrite the Hamiltonian
(\ref{H_0_fermionic})--(\ref{backscattering_fermionic}) in terms of
$\varphi_L$ and $\varphi_R$. It will be more convenient, however, to
present the Hamiltonian in terms of slightly different bosonic fields:
\begin{equation}
\phi_L = \frac{1}{2}(\varphi_{L2} - \varphi_{L1}),
\quad
\Pi_L = -\frac{1}{2\pi} \nabla (\varphi_{L1} + \varphi_{L2}).
  \label{good_fields}
\end{equation}
The new variables satisfy the commutation relations typical for the
displacement of a 1D elastic medium and its momentum density,
\begin{equation}
[\phi_L(x), \Pi_L(y)] = i \delta(x-y).
  \label{good_commutator}
\end{equation}
The form of the free-electron Hamiltonian (\ref{H_0_fermionic}) supports
this observation,
\begin{eqnarray}
H_0 = \frac{\hbar v_F}{2}\int
      &&\left\{\frac{1}{\pi}[\nabla \phi_L(x)]^2 + \pi \Pi_L^2(x)\right.
\nonumber\\
      &&\hspace{.7em}
        \left. +\frac{1}{\pi}[\nabla \phi_R(x)]^2 + \pi \Pi_R^2(x)
        \right\}dx.
  \label{H_0_bosonized}
\end{eqnarray}

This qualitative interpretation allows one to predict the correct
bosonized form of the charging energy (\ref{H_C}). One should expect that
the charge of the dot is proportional to the difference of the charge
$Q_L\sim e\phi_L(0)$ brought into the dot through the left contact and the
charge $Q_R\sim e\phi_R(0)$ carried away through the right contact.
Explicit calculation gives
\begin{equation}
H_C = \frac{E_C}{\pi^2}
      \left[\phi_R(0) - \phi_L(0) -\pi N\right]^2.
  \label{H_C_bosonized}
\end{equation}
It is important to stress that the interaction term (\ref{H_C_bosonized})
is now quadratic in $\phi_L$ and $\phi_R$. Therefore the bosonization
approach enables one to treat the charging energy exactly.

Finally, the backscattering term (\ref{backscattering_fermionic}) takes
the form
\begin{equation}
H' = \frac{D |r_L|}{\pi} \cos[2\phi_L(0)]
     +\frac{D |r_R|}{\pi} \cos[2\phi_R(0)].
  \label{backscattering_bosonized}
\end{equation}
In Eq.~(\ref{backscattering_bosonized}) we neglected the products of
Majorana fermions $\eta_{L1}^{}\eta_{L2}^{}$ and
$\eta_{R1}^{}\eta_{R2}^{}$, as they commute with each other.  Unlike the
first two terms, $H_0$ and $H_C$, the backscattering term is not quadratic
in bosonic variables and cannot be treated exactly. Below we assume weak
backscattering and develop the perturbation theory in small parameters
$|r_L|$ and $|r_R|$.

\subsection{Perturbation theory for the conductance}

In this paper we are only interested in the linear conductance through the
quantum dot. To find it, we will use a Kubo formula, which in the
zero-frequency limit can be written as
\begin{equation}
G = \frac{i}{2\hbar}
    \int_{-\infty}^{\infty} t
    \langle [ I(t), I(0) ] \rangle dt,
  \label{Kubo}
\end{equation}
where $I$ is the operator of current through the dot.

We define $I$ as the average of the current $I_L$ flowing into the dot
through the left contact and the current $I_R$ flowing out of the dot
through the right contact.  Clearly, the current $I_L$ is proportional to
the time derivative of the displacement of the effective elastic medium
describing the left contact, $I_L\sim e\dot\phi_L$, and, similarly,
$I_R\sim e\dot\phi_R$. From the equations of motion corresponding to the
Hamiltonian $H_0 + H_C + H'$, one concludes that $I\sim
ev_F(\Pi_L+\Pi_R)$. A careful calculation gives
\begin{equation}
I = \frac{1}{2}ev_F
      \left.\left(\Pi_L + \Pi_R\right)\right|_{x=0}.
  \label{current_operator}
\end{equation}

As the first step, we will find the conductance in the absence of the
barriers. Since the current operator (\ref{current_operator}) commutes
with $H_C$, one should expect the conductance to be insensitive to the
charging energy. Clearly, at $E_C = 0$ the system is equivalent to two
resistances $2\pi\hbar/e^2$ of the two quantum point contacts, connected
in series. One then expects the conductance to be
\begin{equation}
G_0 = \frac{e^2}{4\pi\hbar}.
  \label{G_0}
\end{equation}
This result is easily derived from the Kubo formula (\ref{Kubo}), because
with $H'=0$ the current-current correlator must be found for a quadratic
Hamiltonian.  As a result, one finds $\langle [ I(t), I(0) ] \rangle =
i(e^2/2\pi)\delta'(t)$, and after the substitution into (\ref{Kubo}) one
reproduces (\ref{G_0}).

The result (\ref{G_0}) is independent of temperature. Since we used the
Hamiltonian (\ref{H_0_fermionic}) with linearized spectra of electrons,
the temperature was assumed to be much smaller than the Fermi energy, but
could still be of the order of the charging energy. In the presence of the
barriers the conductance must acquire some temperature dependence, which
is obvious from the limit of very high barriers (\ref{cotunneling}).
Indeed, the lowest non-vanishing (2nd-order) correction in $r_L$ and $r_R$
to the conductance is already temperature dependent. At low temperatures
$T\ll E_C$, we get
\begin{equation}
G=\frac{e^2}{4\pi\hbar}\!\left(1-\frac{\pi\Gamma_0(N)}{4T}\right)\!,
  \label{2nd_order_spinless}
\end{equation}
where $\Gamma_0$ is defined as follows:
\begin{equation}
\Gamma_0(N)=\frac{2\gamma E_C}{\pi^2}
              \left[|r_L|^2+|r_R|^2+2|r_L||r_R|\cos(2\pi N)\right]\!.
\label{Gamma_0}
\end{equation}
Here $\gamma=e^{\bf C}$, with ${\bf C}\approx 0.5772\ldots$ being the
Euler's constant.  In Eq.~(\ref{2nd_order_spinless}) we have neglected the
terms small in $T/E_C$.  The result (\ref{2nd_order_spinless}) and
(\ref{Gamma_0}) is analogous to that for the conductance of a 1D wire with
two defects\cite{Furusaki,remark3} and can be obtained in a similar way,
see Appendix~\ref{A1}.

As expected, the presence of weak scattering in the contacts gives rise to
a small periodic correction to the conductance. The same behavior was
predicted\cite{Matveev1} for the average charge of the dot. However,
unlike the average charge, the correction to conductance strongly depends
on temperature. As the temperature is lowered, the conductance decreases.
The only case when conductance is temperature-independent is when the
barriers are symmetric, $|r_L|=|r_R|$, and the dimensionless gate voltage
$N$ is half-integer, which corresponds to the center of a peak in the
high-barrier case. In this regime we expect the conductance to be of the
order of $e^2/\hbar$, cf.\ Sec.~\ref{weak}, which is confirmed by the
perturbation theory (\ref{2nd_order_spinless}) and (\ref{Gamma_0}).

\subsection{Non-perturbative calculation}

At low temperatures $T\lesssim \Gamma_0$ the second-order correction
becomes large, and the perturbation theory fails. To find the conductance
in this case, one has to sum up infinite number of terms of the
perturbation theory. To perform such a summation, we will use the
technique developed in Ref.~\onlinecite{Matveev1}.

At the first step we perform a linear transformation of the bosonic
fields,
\begin{equation}
\phi_L = \frac{\phi_I - \phi_C}{\sqrt{2}},
\quad
\phi_R = \frac{\phi_I + \phi_C}{\sqrt{2}},
  \label{transformation}
\end{equation}
and similarly for $\Pi_L$ and $\Pi_R$. The advantage of using these new
variables is that the Coulomb energy (\ref{H_C_bosonized}) is now
expressed in terms of the charging mode $\phi_C$ only,
\begin{equation}
H_C = \frac{2E_C}{\pi^2}
      \left[\phi_C(0)-\frac{\pi N}{\sqrt{2}}\right]^2.
  \label{H_C_diagonalized}
\end{equation}
The charging energy (\ref{H_C_diagonalized}) suppresses the fluctuations
of $\phi_C(0)$ at energy scales below $E_C$. Thus at $T\ll E_C$ one can
integrate out the fluctuations of the charging mode by replacing the
scattering term (\ref{backscattering_bosonized}) with its value averaged
over the fluctuations of $\phi_C$. The resulting Hamiltonian is expressed
in terms of the field $\phi_I$ only and has the form
\begin{eqnarray}
H_0 &=& \frac{\hbar v_F}{2}\int
        \left\{\frac{1}{\pi}[\nabla \phi_I(x)]^2 + \pi \Pi_I^2(x)
        \right\}dx,
  \label{H_0_simplified}\\
H' &=& \left(\frac{\gamma E_C D}{2\pi^3}\right)^{1/2}
       \left[re^{i\sqrt{2}\phi_I(0)}
       + r^{*}e^{-i\sqrt{2}\phi_I(0)}\right],
  \label{H'_simplified}
\end{eqnarray}
where the dependence on the gate voltage is incorporated into the
complex parameter $r = |r_L|e^{-i\pi N} + |r_R|e^{i\pi N}$.

The Hamiltonian (\ref{H_0_simplified}) and (\ref{H'_simplified}) can be
transformed to that of an impurity in a $g=\frac12$ Luttinger
liquid\cite{Kane} by a simple transformation $\phi=\sqrt{2}\phi_I$ and
$\Pi=\Pi_I/\sqrt{2}$. The latter can be solved exactly\cite{Kane} using
the technique developed in Ref.~\onlinecite{Guinea}. For our purposes it
will be more convenient to use an alternative exact
solution.\cite{Matveev1} The idea is to interpret $e^{i\sqrt{2}\phi_I}$ as
a bosonized form of some fermion annihilation operator.\cite{Luther} This
can be done by rewriting the Hamiltonian in terms of two decoupled chiral
bosonic fields defined as
\begin{equation}
\varphi_{\pm}(x) = \frac{\phi_I(x) \pm \phi_I(-x)}{\sqrt{2}}
                   +\pi\int_{0}^{x}
                    \frac{\Pi_I(x') \pm \Pi_I(-x')}{\sqrt{2}} dx'.
  \label{chiralization}
\end{equation}
In these variables the Hamiltonian takes the form
\begin{eqnarray}
H_0 &=& \frac{\hbar v_F}{4\pi}\int
        \left\{[\nabla \varphi_{+}(x)]^2 + [\nabla \varphi_{-}(x)]^2
        \right\}dx,
  \label{H_0_chiralized}\\
H' &=& \left(\frac{\gamma E_C D}{2\pi^3}\right)^{1/2}
       \left[re^{i\varphi_{+}(0)}
       + r^{*}e^{-i\varphi_{+}(0)}\right].
  \label{H'_chiralized}
\end{eqnarray}
One can easily check that the fields $\varphi_+$ and $\varphi_-$ commute.
Thus the $\varphi_-$ part of the Hamiltonian completely decouples.  Since
the field $\varphi_+(x)$ satisfies the same commutation relations
(\ref{commutators}) as $\varphi_{L1}$, one can in complete analogy with
Eq.~(\ref{bosonization}) interpret $e^{i\varphi_+}$ as a fermion
operator,\cite{remark}

\begin{figure}
\narrowtext
\epsfxsize=3in
\hspace{.2em}
\epsffile{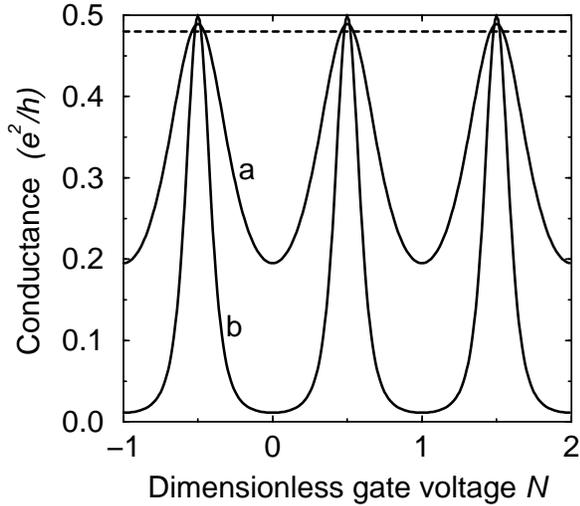}
\vspace{1.5\baselineskip}
\caption{Conductance (\protect\ref{conductance_spinless}) as a function
  of the dimensionless gate voltage $N$ for the symmetric case,
  $|r_L|=|r_R|=0.2$.  The two curves are calculated for $E_C/T=25$ (a) and
  $200$ (b). In this calculation we included the correction proportional
  to $T/E_C$ in Eq.~(\protect\ref{finally_obtain}). The dashed line
  represents the conductance $G=G_LG_R/(G_L + G_R)$ in the
  high-temperature limit, $T\gg E_C$.}
\label{fig:2}
\end{figure}
\vspace{\baselineskip}

\begin{equation}
e^{i\varphi_+(x)} = \left(\frac{2\pi\hbar v_F}{D}\right)^{1/2}
                    \eta_{+}\psi_{+}(x).
  \label{fermionization}
\end{equation}

The resulting Hamiltonian is quadratic in fermion operators,
\begin{eqnarray}
H &=& i\hbar v_F\int\psi_{+}^{\dagger}(x)\nabla\psi_{+}(x)dx
\nonumber\\
  & & + \left(\frac{\gamma \hbar v_F E_C}{\pi^2}\right)^{1/2}
      \left[r\eta_{+}^{}\psi_{+}(0)
            + r^{*}\psi_{+}^{\dagger}(0)\eta_{+}^{}
      \right],
  \label{H_fermionized}
\end{eqnarray}
and can be easily diagonalized.\cite{Matveev1} This greatly simplifies the
calculation of the conductance. One can again use the Kubo formula
(\ref{Kubo}), with the properly transformed current operator
(\ref{current_operator}). In the new fermionic variables the current
operator has the form
\begin{equation}
I = -e v_F \left. :\!\psi_{+}^{\dagger}\psi_{+}\!\!:
           \right|_{x=0}.
  \label{current_operator_fermionized}
\end{equation}

The actual calculation of the linear conductance is rather long, but
straightforward. First we find the current-current correlation function
to be
\begin{eqnarray}
& &\langle [I(t),I(0)]\rangle =
   i\frac{e^2}{2\pi}\delta'(t)
\nonumber\\
 & &\hspace{3em}
  - i\frac{e^2}{2\pi^2\hbar^2}\int\!\!\!\int\!
  \frac{\Gamma_0^2\sin[(\epsilon+\epsilon')t/\hbar]d\epsilon d\epsilon'}
       {(\epsilon^2+\Gamma_0^2)(e^{\epsilon/T}+1)(e^{\epsilon'/T}+1)}.
  \label{current-current}
\end{eqnarray}
The substitution of Eq.~(\ref{current-current}) into the Kubo formula
(\ref{Kubo}) gives the following expression for the linear conductance:
\begin{equation}
G=\frac{e^2}{4\pi\hbar}
  \left[1-\int^\infty_{-\infty}dE
        \frac{1}{4T\cosh^2\frac{E}{2T}}
        \frac{\Gamma^2_0(N)}{E^2+\Gamma^2_0(N)}
  \right].
  \label{conductance_spinless}
\end{equation}
Equation (\ref{conductance_spinless}) is the central result of this
section.  It is valid for any temperature below $E_C$, and, unlike the
perturbative result (\ref{2nd_order_spinless}), it is not restricted to
only sufficiently high temperatures $T\gg\Gamma_0$.  Of course, at
$T\gg\Gamma_0$ the result (\ref{2nd_order_spinless}) is reproduced.

In the opposite limit, $T\ll\Gamma_0$, the conductance is quadratic in
temperature,
\begin{equation}
G = \frac{\pi e^2}{12\hbar}\left[\frac{T}{\Gamma_0(N)}\right]^2.
  \label{low_T_spinless}
\end{equation}
It is important to note that the same temperature dependence was found for
the inelastic co-tunneling (\ref{cotunneling}) in the weak-tunneling
regime. As expected, the quadratic temperature dependence is a universal
property of inelastic co-tunneling and is not limited to the
weak-tunneling limit, cf.\ Eq.~(\ref{tails}).

In the special case of symmetric barriers, $|r_L| = |r_R| = r_0$, the
low-temperature limit (\ref{low_T_spinless}) is never achieved on
resonance. Indeed, we have $\Gamma_0= (8\gamma E_C/\pi^2)r_0^2\cos^2 \pi
N$, i.e., $\Gamma_0$ vanishes at half-integer values of $N$. Therefore in
the low-temperature limit the resonant value of conductance is
$e^2/4\pi\hbar$, whereas away from the resonance $\Gamma_0\neq 0$, and
$G\propto T^2$. It is clear then that at low temperatures the weak Coulomb
blockade oscillation in conductance (\ref{2nd_order_spinless}) transforms
into sharp periodic peaks, Fig.~\ref{fig:2}. Unlike the weak-tunneling
case (\ref{cotunneling}), the tails of the peak at $N=\frac12$ are not
Lorentzian: $G\propto1/(N-\frac12)^4$.

Finally, it is worth noting that the conductance
(\ref{conductance_spinless}) coincides with that of a weak impurity in the
$g=\frac12$ Luttinger liquid.\cite{Kane} This is not surprising,
considering the aforementioned identity of the Hamiltonian
(\ref{H_0_simplified}), (\ref{H'_simplified}) with the one for the quantum
impurity problem.\cite{Kane} In the latter problem the condition
$g=\frac12$ requires special fine-tuning of the strength of interaction
between electrons. In our case the condition $g=\frac12$ is satisfied
automatically when we integrate out one of the two bosonic
modes.\cite{Furusaki}

\section{Co-tunneling of electrons with spins}
\label{withspin}

In the previous section we considered the case of spinless electrons,
which can be realized by applying strong in-plane magnetic field. In the
absence of the field, each of the contacts has two identical modes,
corresponding to two possible values of electron spin. The number of modes
is an important parameter in our problem. This can be easily seen from the
analogy to the Kondo problem discussed in Sec.~\ref{weak}. In the
weak-tunneling case, the model of spinless electrons maps to the 2-channel
Kondo problem, because the fictitious spin of the impurity---i.e., the
charge of the dot---can be changed by tunneling of an electron in two
independent channels, corresponding to the left and right contacts. In the
presence of real spins of electrons, the number of channels in the
effective Kondo problem is 4. It is well known that the number of channels
in the Kondo problem strongly affects its properties.\cite{NB} Thus one
should expect that the presence of spins should have a dramatic effect
upon the transport through a quantum dot.

\subsection{Strong-tunneling case}

We will start with the discussion of the strong-tunneling limit, ${\cal
  T}_L, {\cal T}_R \to 1$. To find the conductance, one can use the same
technique as in Sec.~\ref{spinless}. The Hamiltonian of the problem can be
written in the bosonized form, analogous to
(\ref{H_0_bosonized})--(\ref{backscattering_bosonized}),
\begin{eqnarray}
H_0 &=& \frac{\hbar v_F}{2}\!\sum_{\alpha}\int\!
      \left\{\frac{1}{\pi}[\nabla \phi_{L\alpha}(x)]^2
                           + \pi \Pi_{L\alpha}^2(x)\right.
\nonumber\\
     &&\hspace{5.3em}\left. +\frac{1}{\pi}[\nabla \phi_{R\alpha}(x)]^2
                                         + \pi \Pi_{R\alpha}^2(x)
        \right\}\!dx,
\label{H_0_with_spins}\\
H_{C} &=& \frac{E_C}{\pi^2}\left\{\sum_{\alpha}
         [\phi_{R\alpha}(0) - \phi_{L\alpha}(0)] -\pi N\right\}^2,
\label{H_C_with_spins}\\
H' &=& \frac{D}{\pi} \sum_{\alpha}
       \{|r_L| \cos[2\phi_{L\alpha}(0)]\!
              +\! |r_R| \cos[2\phi_{R\alpha}(0)]
       \}.
  \label{backscattering_with_spins}
\end{eqnarray}
Here the fields with $\alpha=\ \uparrow$ and $\downarrow$ describe the
electrons with spins up and down respectively.

One can again find the linear conductance using the Kubo formula
(\ref{Kubo}), but the current operator (\ref{current_operator}) must be
modified to account for the spins,
\begin{equation}
I = \frac{1}{2}ev_F\!\sum_{\alpha=\uparrow,\downarrow}\!
      \left(\Pi_{L\alpha} + \Pi_{R\alpha}\right)\Bigl|_{x=0}.
  \label{current_operator_with_spins}
\end{equation}

In the absence of barriers, $|r_L|=|r_R|=0$, the calculation is trivial,
as the Hamiltonian (\ref{H_0_with_spins}) and (\ref{H_C_with_spins})
is quadratic. Due to the doubling of the number of modes in each contact,
in the presence of spins the conductance doubles compared to
Eq.~(\ref{G_0}),
\begin{equation}
G_0 = \frac{e^2}{2\pi\hbar}.
  \label{G_0_with_spins}
\end{equation}

Weak backscattering, $|r_L|,|r_R|\ll1$, can be treated in perturbation
theory in $H'$, see Appendix \ref{A2}. Up to the second order we get
\begin{equation}
  G = \frac{e^2}{2\pi\hbar}
     \left[1-\frac{2\Gamma(\frac{3}{4})}{\Gamma(\frac{1}{4})}
             \sqrt{\frac{\gamma E_C}{\pi T}}
             \left(|r_L|^2+|r_R|^2\right)
     \right].
  \label{2nd_order_with_spins}
\end{equation}
It is instructive to compare this result with its analog for the spinless
case, namely Eq.~(\ref{2nd_order_spinless}). In both cases the small
correction to the conductance has the tendency to grow as the temperature
is decreased. An important difference is that the second-order correction
in (\ref{2nd_order_with_spins}) does not depend on the gate voltage $N$,
and as a result it grows even at half-integer $N$, i.e., on resonance.
Thus, in contrast to the spinless case, one should expect the resonance
value to be smaller than the unperturbed conductance $G_0=e^2/2\pi\hbar$.

This difference has a natural interpretation in terms of the analogy with
the Kondo model. As we already mentioned, the spinless case corresponds to
the 2-channel Kondo model. In the 2-channel case the strong-coupling fixed
point is stable at zero magnetic field, or, in terms of the Coulomb
blockade problem, at half-integer $N$. Therefore, precisely on resonance a
weak reflection must be an irrelevant perturbation, and the peak value of
the conductance should be given by Eq.~(\ref{G_0}). On the contrary, the
4-channel Kondo problem corresponding to the case with spins has a stable
fixed point at an intermediate value of coupling, and both weak- and
strong-coupling fixed points are unstable.\cite{NB} Thus even on resonance
the conductance is not given by its value in the strong-tunneling limit
(\ref{G_0_with_spins}), as confirmed by the perturbation theory
(\ref{2nd_order_with_spins}).

\subsection{Conductance in the case of asymmetric barriers}
\label{asymmetric}

Unfortunately, it is not easy to calculate the conductance near the
intermediate-coupling fixed point. Some conclusions about it can be drawn
on the basis of the scaling approach, Sec.~\ref{scaling}. One should note,
however, that this fixed point is stable only if the coupling in all the 4
channels is identical. In the case of the Coulomb blockade problem, the
coupling in the pairs of channels corresponding to different spin
orientations in the same contact must be identical (in the absence of
magnetic field). On the other hand, the coupling constants for the modes
in different contacts are determined by the corresponding conductances and
are not necessarily equal. In fact, in the structure shown in
Fig.~\ref{fig:1}, the two conductances are controlled separately by the
gate voltages $V_L$ and $V_R$. Therefore in experiments the coupling
constants corresponding to the left and right modes are not equal, unless
a special attempt to make them equal is made.

In this section we study the limit of very asymmetric barriers, namely we
assume that one of the barriers is very high, whereas the other one is
low. In terms of the transmission coefficients of the barriers this limit
corresponds to ${\cal T}_L \ll 1$ and $1-{\cal T}_R \ll 1$. Such a regime
can be easily realized in an experiment by proper tuning of the gate
voltages $V_L$ and $V_R$. In addition, we will see in Sec.~\ref{scaling}
that at $T\to 0$ arbitrarily weak asymmetry grows and the system scales
towards the strongly asymmetric limit. Therefore at low temperatures the
results of this section can be applied to any asymmetric system.

In Sec.~\ref{weak} we described the contact in the regime of weak
tunneling by the tunnel Hamiltonian (\ref{tunnel_hamiltonian}) written in
terms of the original electrons. On the other hand, in Sec.~\ref{spinless}
it was more convenient to use the bosonized form of the Hamiltonian for
the electron gas in the contact with transmission coefficient close to
unity. Therefore when considering the asymmetric case ${\cal T}_L\to0$ and
${\cal T}_R\to1$ we shall bosonize only the electrons in the right
contact. Then the unperturbed fixed-point Hamiltonian $H_0 + H_C$ is given
by
\begin{eqnarray}
H_0 &=& \sum_{k\alpha} \epsilon_k a_{k\alpha}^\dagger a_{k\alpha}^{}
        + \sum_{p\alpha} \epsilon_p a_{p\alpha}^\dagger a_{p\alpha}^{}
\nonumber\\
    & & +\frac{\hbar v_F}{2}\!\sum_{\alpha}\!\int\!
         \left\{\frac{1}{\pi}[\nabla \phi_{R\alpha}(x)]^2
                                         + \pi \Pi_{R\alpha}^2(x)
        \right\}dx,
  \label{H_0_asymmetric}\\
H_C &=& E_C\!
        \left[\hat n+\frac{1}{\pi}\sum_\alpha\phi_{R\alpha}(0)
              - N\right]^2.
  \label{H_C_asymmetric}
\end{eqnarray}
Here the operators $a_{k\alpha}$ and $a_{p\alpha}$ correspond to the
left-contact electrons in the lead and in the dot, respectively; the
bosonic fields $\phi_{R\alpha}$ and $\Pi_{R\alpha}$ model the electrons in
the right contact. An integer-valued operator $\hat n$ describes the number
of electrons transferred to the dot through the left barrier.

The perturbation consists of two parts: weak tunneling through the high
barrier in the left contact, and weak scattering on the small barrier in
the right contact,
\begin{eqnarray}
  H_L' &=& \sum_{kp\alpha}
           (v_t^{} a_{k\alpha}^\dagger a_{p\alpha}^{}F
            + v_t^* a_{p\alpha}^\dagger a_{k\alpha}^{}F^\dagger),
\label{H_L'}\\
  H_R' &=& \frac{1}{\pi}D|r_R| \sum_{\alpha}
           \cos[2\phi_{R\alpha}(0)].
\label{H_R'}
\end{eqnarray}
We have chosen not to write the operator $\hat n$ of the number of
electrons transferred into the dot through the barrier in terms of the
fermionic operators as $\hat n=\sum a_{p\alpha}^\dagger a_{p\alpha}^{}$,
but to treat it as an independent variable. Thus the tunneling Hamiltonian
(\ref{H_L'}) acquires the charge-lowering and raising operators $F$ and
$F^\dagger$ defined as
\begin{equation}
[F,\hat n] = F.
  \label{lowering_operator}
\end{equation}
One possible representation\cite{Ingold} of these operators is given by
$F=e^{ix}$ and $\hat n = i\frac{\partial}{\partial x}$, where $x$ has the
meaning of a phase operator conjugated to $\hat n$.

Our goal in this section is to find the conductance of the system,
assuming that the perturbations $H_L'$ and $H_R'$ are small. At the first
step we will utilize the smallness of the amplitude $v_t^{}$ of tunneling
through the left barrier, and find the expression for the current in the
second order in $v_t^{}$. The current can be defined as the time
derivative of the number of electrons that have tunneled through the
barrier, $I = -e \frac{\partial \hat n}{\partial t} =
\frac{ie}{\hbar}[\hat n,H]$. Using the commutation relation
(\ref{lowering_operator}), we get
\begin{equation}
I = i\frac{e}{\hbar}
    \sum_{kp\alpha} \left(
    v_t^* a_{p\alpha}^\dagger a_{k\alpha}^{} F^\dagger -
    v_t a_{k\alpha}^\dagger a_{p\alpha}^{} F\right).
  \label{current_asymmetric}
\end{equation}
To find the conductance through the dot, we will substitute the above
current operator into the Kubo formula (\ref{Kubo}). This requires the
calculation of the correlator $\langle[I(t),I(0)]\rangle$. Since $I\propto
|v_t|$, up to the second order in tunneling matrix elements we can average
$[I(t),I(0)]$ over the equilibrium thermal distribution of the system
without tunneling, $H_L'=0$. Then the averaging over the fermionic degrees
of freedom is straightforward, and we get
\begin{equation}
G = -i G_L \frac{\pi T^2}{\hbar^2}
      \int_{-\infty}^{\infty}
      \frac{tK(t)dt}{\sinh^2[\pi T(t-i\delta)/\hbar]}.
  \label{conductance_asymmetric}
\end{equation}
Here we have introduced the conductance of the left tunnel barrier,
$G_L=(4\pi e^2/\hbar)
\sum_{kp}|v_t^{}|^2\delta(\epsilon_k)\delta(\epsilon_p)$, and the
correlator
\begin{equation}
K(t) = \langle F(t) F^\dagger(0)\rangle
        = \langle F^\dagger(t) F(0)\rangle.
  \label{K}
\end{equation}
The equality of the two correlators in Eq.~(\ref{K}) follows from the fact
that the symmetry transformation $\hat n \to -\hat n$, $\phi_{R\alpha}\to
-\phi_{R\alpha}$, and $N\to -N$ does not affect the Hamiltonian
(\ref{H_0_asymmetric}), (\ref{H_C_asymmetric}), and (\ref{H_R'}), but
changes $F^\dagger\leftrightarrow F$.

It is instructive to apply the formula (\ref{conductance_asymmetric}) to
the non-interacting case, $E_C = 0$. In this limit the operators $F$ and
$F^\dagger$ commute with the Hamiltonian (\ref{H_0_asymmetric}) and
(\ref{H_R'}). Thus the correlator (\ref{K}) is unity, and we get $G=G_L$.
This is the expected result, since we assumed $G_L\to0$, and the
conductance of the system must be determined by that of the weakest link.

Let us now find the conductance at non-zero charging energy $E_C$, but in
the absence of the scattering in the right contact, $r_R=0$. In this
case the correlator (\ref{K}) is no longer trivial, because the operators
$F$ and $F^\dagger$ do not commute with the interaction term
(\ref{H_C_asymmetric}). When an electron tunnels into the dot, $n\to n+1$,
the system is shifted from its equilibrium state.  To accommodate the
additional electron brought by the tunneling process and to return the
system to the equilibrium state, one electron must leave the dot
through the right contact. Thus the time evolution of the operators $F$
and $F^\dagger$ becomes non-trivial.

To find the correlator $K(t)$, we will perform the following unitary
transformation of the Hamiltonian:
\begin{eqnarray}
&U = e^{i\hat n \Theta},&
  \label{unitary_transformation}\\
&{\displaystyle
\Theta = \frac{\pi}{2} \int_{-\infty}^{\infty}
            \left[\Pi_{R\uparrow}(y) + \Pi_{R\downarrow}(y)\right]dy}.&
  \label{phase}
\end{eqnarray}
This transformation shifts the bosonic fields $\phi_{R\alpha}$ and adds a
phase factor to the charge-lowering operator $F$,
\begin{eqnarray}
 \phi_{R\alpha}(x)&\to&  \phi_{R\alpha}(x) - \frac{\pi}{2}\hat n,
\label{transformed_phi}\\
 F&\to& F e^{i\Theta}.
\label{transformed_F}
\end{eqnarray}
One can easily see that the kinetic-energy term (\ref{H_0_asymmetric}) is
invariant under the transformation (\ref{unitary_transformation}), whereas
the charging term (\ref{H_C_asymmetric}) transforms to a $\hat
n$-independent form,
\begin{equation}
\tilde H_C = U^\dagger H_C U =
E_C\left[\frac{1}{\pi}\sum_\alpha\phi_{R\alpha}(0) - N\right]^2.
\label{transformed_H_C}
\end{equation}
Upon this transformation, the operators $F$ and $F^\dagger$ commute with
the Hamiltonian (\ref{H_0_asymmetric}) and (\ref{transformed_H_C}), but
they acquire non-trivial phase factors, see Eq.~(\ref{transformed_F}).
This enables us to find $K(t)$ as the correlator of the phase factors,
$K(t) = K_{\Theta}(t)\equiv \langle e^{i\Theta(t)}
e^{-i\Theta(0)}\rangle$.  Since the phase $\Theta$ is linear in bosonic
variables, the calculation of the correlator $K_{\Theta}$ is
straightforward:
\begin{eqnarray}
K_{\Theta}(t)
   &=& \exp\{-\langle[\Theta(0)-\Theta(t)]\Theta(0)\rangle\}
\nonumber\\
   &\simeq& \frac{\pi^2 T}{2i\gamma E_C}
         \frac{1}{\sinh[\pi T(t-i\delta)/\hbar]}.
  \label{Debye_Waller}
\end{eqnarray}
The last equality is valid asymptotically at low energies, $T,
\hbar/t\ll E_C$. The substitution of $K=K_\Theta$ into the
expression for the conductance (\ref{conductance_asymmetric}) leads to a
linear temperature dependence of the conductance:
\begin{equation}
G = G_L \frac{\pi^3 T}{8\gamma E_C}.
  \label{peak_conductance_asymmetric}
\end{equation}

Physically the origin of the linear suppression of the conductance at low
temperatures (\ref{peak_conductance_asymmetric}) is the orthogonality
catastrophe.\cite{Anderson2} Unlike the non-interacting case $E_C = 0$
discussed above, the charge fluctuations in the dot are now suppressed at
energies below $E_C$. Effectively one can interpret the effect of the
charging energy as a hard-wall boundary condition for the wavefunctions of
the right-contact electrons with energies within the band of width $W\sim
E_C$ near the Fermi level. When an electron tunnels through the left
barrier, the system must lower its charging energy by moving one electron
through the right contact. Since the two spin channels are completely
symmetric, each mode transfers the charge $q=\frac{e}{2}$. According to
the Friedel sum rule, this leads to a change in the boundary condition
corresponding to an additional scattering phase shift $\delta=\pm\frac{\pi
  q}{e}=\pm\frac{\pi}{2}$.  A sudden change of the boundary conditions
creates a state with a large number of electron-hole excitations, which is
nearly orthogonal\cite{Anderson2} to the ground state of the system. This
orthogonality leads to a power-law suppression of the tunneling density of
states $\nu(\varepsilon)\propto\varepsilon^\chi$, with the
exponent\cite{Schotte,ND,Mahan} determined by the phase shifts in all the
electronic channels, $\chi=\sum(\delta/\pi)^2$. In our system, each spin
mode gives two channels---in the dot and in the right lead---leading to
the total of 4 channels. Thus the exponent $\chi$ is unity, and the
density of states $\nu(\varepsilon)\propto\varepsilon$. The linear
suppression of the tunneling density of states results in the $T$-linear
conductance (\ref{peak_conductance_asymmetric}).

The linear temperature dependence of conductance
(\ref{peak_conductance_asymmetric}) appears to contradict to the $T^2$ law
for the inelastic co-tunneling, Eqs.~(\ref{cotunneling}), (\ref{tails})
and (\ref{low_T_spinless}). The reason is the absence of the barrier in
the right junction assumed in the derivation of
Eq.~(\ref{peak_conductance_asymmetric}). We will now show that the
presence of arbitrarily weak barrier in the right contact is crucial (see
also Appendix~\ref{A3}) and leads to the quadratic temperature dependence
of the conductance at $T\to0$.

The weak backscattering in the right contact is described by the term
(\ref{H_R'}) in the Hamiltonian.  As it follows from
Eq.~(\ref{transformed_phi}), under the unitary transformation
(\ref{unitary_transformation}) the scattering term (\ref{H_R'}) acquires
additional sign:
\begin{equation}
\tilde H_R' = (-1)^{\hat n}
              \frac{1}{\pi}D|r_R| \sum_{\alpha}
              \cos[2\phi_{R\alpha}(0)].
  \label{transformed_H_R'}
\end{equation}
The operators $F$ and $F^\dagger$ no longer commute with the Hamiltonian:
they commute with $H_0 + \tilde H_C$ and anti-commute with $\tilde H_R'$.
Therefore the correlator (\ref{K}) is no longer equal to $K_{\Theta}$, but
rather $K(t)=K_{\Theta}(t)K_{F}(t)$, where
\begin{equation}
K_F(t) = \langle F(t) F^\dagger(0)\rangle
  \label{K_F}.
\end{equation}
Here the averaging $\langle\ldots\rangle$ is over the equilibrium thermal
distribution of the transformed Hamiltonian, $H=H_0+\tilde H_C + \tilde
H_R'$.

In the absence of the backscattering in the right contact, $K_{\Theta}$ is
given by Eq.~(\ref{Debye_Waller}), and $K_F(t)\equiv 1$. When a weak
backscattering is added, both correlators are affected. It is clear,
however, that $K_\Theta$ is not modified significantly. Indeed, it follows
from the above mentioned analogy with the orthogonality catastrophe that
the time dependence (\ref{Debye_Waller}) is determined solely by the
charge transferred through the right junction, which in our case is always
$e$---the charge of the electron brought into the dot through the left
barrier---and is not affected by the presence of the right barrier.
Therefore the only change in $K_\Theta(t)$ caused by the weak barrier can
be a small modification of the prefactor in Eq.~(\ref{Debye_Waller}). On
the other hand, we will show that $K_F(t)$ is modified significantly,
which leads to the quadratic temperature dependence of the conductance.

To find $K_F(t)$ we first note that the Hamiltonian $H=H_0+\tilde H_C +
\tilde H_R'$ is identical to the one considered in Sec.~\ref{spinless} and
in Ref.~\onlinecite{Matveev1}. The difference with the problem considered
in Sec.~\ref{spinless} is that instead of two modes corresponding to two
different contacts we now have two spin modes in the same (right) contact.
Thus the two scattering amplitudes are equal and given by $|r_R|$.
As we have seen in Sec.~\ref{spinless}, the presence of the backscattering
shows up at low-energy scales $\epsilon\sim |r_R|^2 E_C\ll E_C$, where one
can effectively integrate out the mode related to the charge fluctuations
in the dot, $\phi_{R\uparrow} + \phi_{R\downarrow}$. The resulting
Hamiltonian is expressed in terms of the spin field
$\phi_{Rs}=(\phi_{R\uparrow} - \phi_{R\downarrow})/\sqrt{2}$ and has the
form:
\begin{eqnarray}
H &=& \frac{\hbar v_F}{2}\!\int\!
         \left\{\frac{1}{\pi}[\nabla \phi_{Rs}(x)]^2
                                         + \pi \Pi_{Rs}^2(x)
        \right\}dx
\nonumber\\
  & & + (-1)^{\hat n}
        \sqrt{\frac{8\gamma E_C D}{\pi^3}}\,|r_R|\cos\pi N
        \cos\!\left[\sqrt{2}\phi_{Rs}(0)\right]\!.
  \label{spin_Hamiltonian}
\end{eqnarray}
We can now fermionize the Hamiltonian using a representation of the
operators $e^{\pm i\sqrt{2}\phi_{Rs}}$ in terms of the fermion creation
and annihilation operators in a way identical to the one leading from the
Hamiltonian (\ref{H_0_simplified}) and (\ref{H'_simplified}) to
Eq.~(\ref{H_fermionized}). The final Hamiltonian is again quadratic in
fermion operators, but also depends on the operator $\hat n$ of the number
of particles transferred through the left barrier,
\begin{eqnarray}
H &=& i\hbar v_F\int\psi^{\dagger}(x)\nabla\psi(x)dx
\nonumber\\
  & & + (-1)^{\hat n}
      \frac{2}{\pi}\sqrt{\gamma \hbar v_F E_C}\,
      |r_R|\cos\pi N
\nonumber\\
  & &\hspace{1em}\times
     (c+c^\dagger)[\psi(0) - \psi^{\dagger}(0)].
  \label{H_fermionized_asymmetric}
\end{eqnarray}
Here we presented a Majorana fermion operator similar to $\eta_+$ in
Eq.~(\ref{H_fermionized}) as a sum of fermion creation and annihilation
operators, $\eta=c+c^\dagger$. This representation is more convenient,
because we can now replace the operators $F$ and $F^\dagger$ in the
definition (\ref{K_F}) of the correlator $K_F$ by
$F_s=F_s^\dagger=1-2c^\dagger c$. Indeed, these new operators commute with
the first term in the Hamiltonian (\ref{H_fermionized_asymmetric}),
anti-commute with the second one, and possess the property
$F_s^{}F_s^\dagger = FF^\dagger=1$. Therefore one can find the correlator
$K_F$ as\cite{remark2}
\begin{equation}
K_F(t) = \langle[1-2c^\dagger(t)c(t)]
              [1-2c^\dagger(0)c(0)]\rangle.
  \label{K_F_fermionized}
\end{equation}

The actual calculation of the correlator (\ref{K_F_fermionized}) is now
straightforward, as the Hamiltonian (\ref{H_fermionized_asymmetric}) is
quadratic in the fermion operators. The result for the correlator $K_F$
has the form
\begin{equation}
K_F(t) = \frac{2\Gamma_R}{\pi}\int_{-\infty}^{\infty}
            \frac{e^{iEt}dE}
                 {(E^2+\Gamma_R^2)(e^{E/T}+1)}.
  \label{K_F_final}
\end{equation}
Here we have introduced the low energy scale
\begin{equation}
\Gamma_R = \frac{8\gamma}{\pi^2}E_C |r_R|^2 \cos^2 \pi N,
  \label{Gamma_R}
\end{equation}
originating from the presence of a weak scatterer in the right contact. In
the weak-scattering limit, $\Gamma_R\to0$, Eq.~(\ref{K_F_final})
reproduces the result $K_F(t)=1$ mentioned above. However, in
the opposite limit $T,\hbar/t\ll\Gamma_R$, the correlator $K_F$ has the
same non-trivial time dependence as the correlator $K_{\Theta}(t)$
given by Eq.~(\ref{Debye_Waller}),
\[
K_F(t) \simeq \frac{2T}{i\Gamma_R}
                 \frac{1}{\sinh[\pi T (t -i\delta)/\hbar]}.
\]
This additional time-dependent contribution to $K(t)$ leads to the
quadratic temperature dependence of the linear conductance at
$T\ll\Gamma_R$.

To find the conductance at temperatures $T\sim\Gamma_R$, we can now
substitute the correlator $K(t)=K_\Theta(t)K_F(t)$ into
Eq.~(\ref{conductance_asymmetric}), which leads to the following
expression:
\begin{equation}
G = \frac{TG_L}{8\gamma E_C}
    \int^\infty_{-\infty}
    \frac{\Gamma_R}{E^2+\Gamma^2_R}
    \frac{\pi^2+(E/T)^2}{\cosh^2(E/2T)}dE.
  \label{conductance_asymmetric_final}
\end{equation}
Equation (\ref{conductance_asymmetric_final}) is the central result of
this section. It is valid for any $T/\Gamma_R$, if both temperature and
$\Gamma_R$ are much smaller than the charging energy $E_C$. The condition
$\Gamma_R\ll E_C$ means $|r_R|\ll1$, i.e., the scattering in the right
contact must indeed be weak.

\begin{figure}
\narrowtext
\epsfxsize=2.8in
\hspace{.4em}
\epsffile{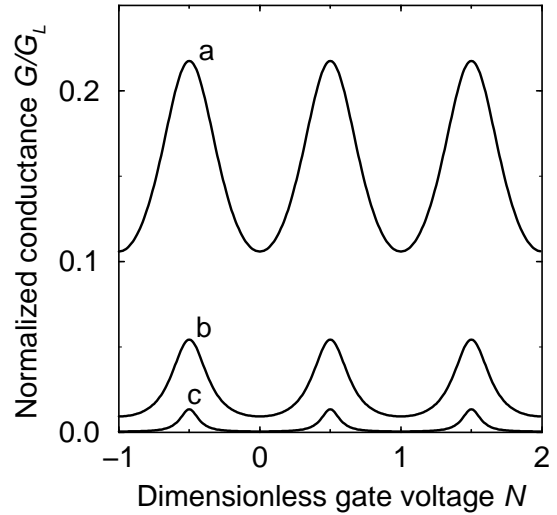}
\vspace{1.5\baselineskip}
\caption{Conductance (\protect\ref{conductance_asymmetric_final})
  as a function of the dimensionless gate voltage $N$ for the asymmetric
  case, $|r_R|=0.4$.  The three curves are calculated for
  $E_C/T=10$ (a), $40$ (b), and $160$ (c).}
\label{fig:3}
\end{figure}
\vspace{\baselineskip}

In the limit $T\gg\Gamma_R$, Eq.~(\ref{conductance_asymmetric_final})
reproduces the previously obtained linear temperature dependence
(\ref{peak_conductance_asymmetric}) in the absence of the scattering in
the right contact. It is important to note, however, that the linear
temperature dependence is obtained not only at $r_R=0$, but also in the
presence of the barrier if $N$ is half-integer, i.e., at the centers of
conductance peaks.

Away from the centers of the peaks, the conductance has a stronger
temperature dependence, $G\propto T^2$. Indeed, at $T\ll\Gamma_R$ the
expression (\ref{conductance_asymmetric_final}) reduces to
\begin{equation}
G = \frac{2\pi^2}{3\gamma}\frac{T^2}{E_C\Gamma_R} G_L.
  \label{off-peak_conductance}
\end{equation}
Thus the quadratic temperature dependence of the linear conductance is
restored for the off-peak values of the gate voltage. The evolution of the
conductance peaks is illustrated in Fig.~\ref{fig:3}.

\section{Scaling approach to inelastic co-tunneling}
\label{scaling}

In the sections \ref{weak}--\ref{withspin} we have considered several
cases when the inelastic co-tunneling can be studied analytically even
though one or both contacts are in the strong-tunneling regime. A common
property of all these results is that at $T\to0$ the conductance behaves
as $T^2$ away from the resonance values of the gate voltage, i.e., at
$N\neq\pm\frac12, \pm\frac32,\ldots$. The same temperature dependence was
obtained earlier\cite{AN} for the off-resonance conductance in the
weak-tunneling case, see Eq.~(\ref{cotunneling}). In this section we study
the general scaling properties of our system and argue that the quadratic
temperature dependence of linear conductance is universal and should hold
for arbitrary barriers.

We will describe our system by the bosonized Hamiltonian
(\ref{H_0_bosonized})--(\ref{backscattering_bosonized}) in the spinless
case and (\ref{H_0_with_spins})--(\ref{backscattering_with_spins}) in the
presence of spins. In the absence of the backscattering the Hamiltonian is
quadratic, i.e., $|r_L|=|r_R|=0$ is a fixed point of our Hamiltonian. In
analogy with the Kondo problem we will call it the strong-coupling fixed
point. To investigate the stability of this fixed point, one should find
the scaling dimension of the backscattering term
(\ref{backscattering_bosonized}) or (\ref{backscattering_with_spins}).

We shall start with the spinless case, where at $T=0$ and $t\gg\hbar/E_C$
a simple calculation gives
\begin{eqnarray*}
\langle H'\!(t)H'\!(0)\rangle &=& \frac{\hbar\Gamma_0(N)}{2\pi i t}
\\
   &&+ \frac{\gamma\hbar^3(|r_L|^2 + |r_R|^2 - 2|r_L||r_R|\cos2\pi N)}
          {4\pi i E_Ct^3},
\end{eqnarray*}
with $\Gamma_0(N)$ defined by Eq.~(\ref{Gamma_0}). The long-time
asymptotics of the correlator $\langle H'(t)H'(0)\rangle$ is given by the
first term, which is proportional to $t^{-1}$, unless $\Gamma_0=0$. Thus
the dimension of the perturbation $H'$ is $\frac12<1$, and the
strong-coupling fixed point is unstable. At a half-integer $N$ and in a
perfectly symmetric case, $|r_L|=|r_R|$, the parameter $\Gamma_0$
vanishes, and the dimension of the operator $H'$ becomes $\frac32>1$. In
this case the strong-coupling fixed point is stable. In the language of
the corresponding 2-channel Kondo problem this means that the
strong-coupling fixed point is stable only in the absence of magnetic
field and channel anisotropy. The leading irrelevant operator with
dimension $\frac32$ gives rise to the weak temperature dependence of the
peak heights in the symmetric spinless case, see Appendix~\ref{A1}.

A similar analysis can be performed in the case of electrons with spins,
described by the Hamiltonian
(\ref{H_0_with_spins})--(\ref{backscattering_with_spins}). The correlator
of the perturbation operators now has the form
\[
\langle H'(t)H'(0)\rangle =
    \bigl(|r_L|^2+|r_R|^2\bigr)
    \sqrt{\frac{4\gamma E_C}{\pi^5}}\,
    \left(\frac{\hbar}{it}\right)^{3/2},
\]
meaning that $H'$ has dimension $\frac34<1$ at any $N$ and any barrier
strengths. Thus as expected the strong-coupling fixed point in the
4-channel case is unstable. At a half-integer $N$ this statement has a
well-known analog in the theory\cite{NB} of the multichannel Kondo
problem, where the strong-coupling fixed point is unstable even in
the absence of the magnetic field.

The fact that the strong-coupling fixed point is unstable means that as
the temperature is lowered, the weak backscattering in the contacts
becomes stronger. It is natural to assume that the effective barriers grow
indefinitely, and the system approaches the weak-coupling fixed point. To
test this hypothesis we will consider a system where the two contacts are
in the weak-tunneling regime and study the stability of the corresponding
fixed point.

As we saw in Sec.~\ref{weak} and \ref{withspin}, in the weak-tunneling
case it is convenient to describe the system in terms of the original
fermionic variables. Generalizing the Hamiltonian of Sec.~\ref{asymmetric}
to the case of both barriers being in the weak-tunneling regime, we
introduce the Hamiltonian $H=H_0 + H_C + H_L'+H_R'$, where
\begin{eqnarray}
H_0 &=& \sum_{k\alpha} \epsilon_k a_{k\alpha}^\dagger a_{k\alpha}^{}
        + \sum_{p\alpha} \epsilon_p a_{p\alpha}^\dagger a_{p\alpha}^{}
\nonumber\\
    & & + \sum_{q\alpha} \epsilon_q a_{q\alpha}^\dagger a_{q\alpha}^{}
        + \sum_{s\alpha} \epsilon_s a_{s\alpha}^\dagger a_{s\alpha}^{} ,
\label{H_0_both_weak}\\
H_C &=& E_C(\hat n_L + \hat n_R - N)^2,
\label{H_C_both_weak}\\
  H_L' &=& \sum_{kp\alpha}
           (v_L^{} a_{k\alpha}^\dagger a_{p\alpha}^{}F_L^{}
            + v_L^* a_{p\alpha}^\dagger a_{k\alpha}^{}F_L^\dagger),
\label{H_L'_both_weak}\\
  H_R' &=& \sum_{qs\alpha}
           (v_R^{} a_{s\alpha}^\dagger a_{q\alpha}^{}F_R^{}
            + v_R^* a_{q\alpha}^\dagger a_{s\alpha}^{}F_R^\dagger).
\label{H_R'_both_weak}
\end{eqnarray}
Here $a_{k\alpha}$ and $a_{s\alpha}$ describe the electrons in the left
and right leads, respectively; $a_{p\alpha}$ and $a_{q\alpha}$ correspond
to the electrons in the dot coupled to the left and right contacts.
Similarly to the formalism of Sec.~\ref{asymmetric}, we have introduced
the operators $\hat n_L$ and $\hat n_R$ of the number of electrons
tunneled through the left and right contacts and the corresponding raising
and lowering operators $F_L^{}$, $F_L^\dagger$, $F_R^{}$, and
$F_R^\dagger$ in the tunnel Hamiltonians (\ref{H_L'_both_weak}) and
(\ref{H_R'_both_weak}).

Away from the resonance values of the gate voltage, $N\neq \pm\frac12,
\pm\frac32,\ldots$, the states with different $n_L$ and $n_R$ have different
energies because of the Coulomb term (\ref{H_C_both_weak}). Since
the tunnel Hamiltonian given by (\ref{H_L'_both_weak}) and
(\ref{H_R'_both_weak}) changes $n_L$ or $n_R$, it does not describe the
transitions between the low-energy states of the Hamiltonian. Thus when
considering the low-energy properties of the system, one should use the
tunneling terms (\ref{H_L'_both_weak}) and (\ref{H_R'_both_weak}) to
construct the perturbation which is not accompanied by the change of the
total charge of the dot. In the simplest case, this can be done by
selecting the terms containing equal number of operators $F_i$ and
$F_i^\dagger$ in the second-order perturbation
\begin{equation}
 H_2' = (H_L' + H_R')\frac{1}{E_0 - H_0 - H_C}(H_L' + H_R').
\label{H_2'}
\end{equation}
Here the denominator accounts for the energy of the virtual state created
by adding (or removing) an electron to the dot. At very low energies the
only important contribution to the denominator is the change of the
electrostatic energy of the system.

At $N$ in the interval $\bigl(-\frac12, \frac12\bigr)$ a typical
representative of the variety of terms contained in Eq.~(\ref{H_2'}) has
the form
\begin{eqnarray}
A &=& -v_L^* v_R^{}\left[\frac{1}{E_C(1-2N)}+\frac{1}{E_C(1+2N)}\right]
\nonumber\\
  & & \times
      F_L^\dagger F_R^{}
      \sum_{kp\alpha}  a_{p\alpha}^\dagger a_{k\alpha}^{}
      \sum_{qs\alpha'}  a_{s\alpha'}^\dagger a_{q\alpha'}^{}.
  \label{A}
\end{eqnarray}
The terms included in Eq.~(\ref{A}) correspond to the processes where one
electron tunnels into the dot through the left junction and another one
escapes from the dot through the right junction. Depending on which of the
two tunneling events is executed first, we get different energies of the
virtual states, corresponding to the two denominators in Eq.~(\ref{A}).

To find out whether or not the weak-coupling fixed point is stable, one
should evaluate the scaling dimension of operator $A$ and other similar
contributions in Eq.~(\ref{H_2'}). A straightforward calculation gives
\begin{eqnarray}
\langle A(t) A^\dagger(0)\rangle &=&
   \frac{\hbar^2G_L G_R}{(2\pi e^2)^2}\!
   \left[\frac{1}{E_C(1-2N)}+\frac{1}{E_C(1+2N)}\right]^2
\nonumber\\
  & &\times
   \left\{\frac{\pi T}{\sinh[\pi T(t-i\delta)/\hbar]}\right\}^4.
  \label{A_correlator}
\end{eqnarray}
At $T\to0$ the correlator decays as $1/t^4$, and therefore
the dimension of the operator $A$ is 2. One can also check that the other
contributions to Eq.~(\ref{H_2'}) have the same dimension, whereas the
higher-order terms have even higher dimensions. Thus the weak-coupling
fixed point is stable.

So far we have shown that away from the resonance values of the gate
voltage, $N\neq \pm\frac12, \pm\frac32, \ldots$, the strong-coupling fixed
point is unstable whereas the weak-coupling fixed point is stable. We now
conjecture that there are no fixed points at intermediate coupling. Then
at low energies the system always evolves towards the stable
(weak-coupling) fixed point. One should also expect that near this fixed
point the temperature dependence of conductance is determined by the
dimension of the leading irrelevant perturbation, which is a universal
characteristic of the problem. Indeed, one can easily show that the
leading contribution to the current operator $I=e\dot n_L$ is given by
\begin{equation}
I = -\frac{ie}{\hbar}(A-A^\dagger).
  \label{current_in_terms_of_A}
\end{equation}
We now substitute this current operator into the Kubo formula
(\ref{Kubo}), and use (\ref{A_correlator}) to find the following
expression for the conductance:
\begin{equation}
G = \frac{\pi\hbar G_LG_R}{3e^2}
   \left[\frac{1}{E_C(1-2N)}+\frac{1}{E_C(1+2N)}\right]^2 T^2.
  \label{cotunneling_conductance}
\end{equation}
This expression for the conductance in the weak-tunneling limit coincides
with the well-known result for the inelastic co-tunneling,\cite{AN} and
reproduces Eq.~(\ref{cotunneling}) near the resonances,
\mbox{$\frac12\pm N \ll 1$}.

It is important to note that in the derivation of
Eq.~(\ref{cotunneling_conductance}) we neglected all the contributions
to the current operator with dimensions $d>2$. Such terms would lead to
the additional terms in the correlator (\ref{A_correlator}), which are
proportional to $\{\pi T/\sinh[\pi T(t-i\delta)/\hbar]\}^m$ with $m>4$.
Clearly, the corresponding corrections to the conductance would behave as
$\delta G \propto T^{m-2}\ll T^2$. Thus at $T\to 0$ the temperature
dependence of the conductance is given by the leading irrelevant
perturbation. As we have seen, the universal scaling dimension $d=2$ of
this perturbation translates into the universal temperature dependence of
the conductance $G\propto T^2$. Therefore this temperature dependence is
specific not only for the cases of weak or strong tunneling, but should
hold for any barrier strengths.

The above treatment of the stability of the weak-coupling fixed point was
limited to the off-resonance values of the gate voltage. We will now
demonstrate that the properties of the system are completely different at
half-integer $N$. Consider, for example, the case of $N=\frac12$. Then the
states with the dot charge equal to 0 and $e$ have the same energy. As a
result the terms in the tunnel Hamiltonians (\ref{H_L'_both_weak}) and
(\ref{H_R'_both_weak}) responsible for the transitions between these two
states no longer lead to the large increase in the energy of the system
and should not be discounted. It is easy to check that the dimension of
such operators is $d=1$, i.e., the tunnel Hamiltonian is a marginal
operator.  This conclusion is actually obvious, because the same
calculation of scaling dimension must be valid in the absence of the
charging effects, $E_C=0$, where we do not expect the tunneling amplitudes
to be renormalized as the temperature is lowered.  However, the simple
calculation of the scaling dimension is not sufficient to determine the
stability of the weak-coupling fixed point. To make such a determination,
one has to study the scaling equations for the tunneling amplitudes
including the higher-order terms in $v_L$ and $v_R$.  We have performed
this analysis in Sec.~\ref{weak}, where we saw from the mapping to the
multichannel Kondo problem that the tunneling amplitude is in fact a
marginally relevant perturbation, i.e., it grows logarithmically at low
temperatures.  Therefore on resonance the weak-coupling fixed point is
unstable.

The on-resonance scaling properties of our system are strongly affected by
the presence of electron spins and by the symmetry of the two barriers.
Let us first consider the case of symmetric barriers.  In the spinless
case we saw that on resonance the strong-coupling fixed point is stable if
the barriers are symmetric. Thus one should expect the system to always
renormalize towards this fixed point. Then the conductance must be on the
order of $e^2/\hbar$. This is confirmed by the explicit calculation in the
strong-tunneling case, Eq.~(\ref{conductance_spinless}).

In the presence of spins, both weak- and strong-coupling fixed points are
unstable on resonance. Thus one should expect that there must be a stable
fixed point at some intermediate coupling strength. This hypothesis is
also confirmed by the analogy with the 4-channel Kondo problem discussed
in Sec.~\ref{weak}. Indeed, it is well-known that the low-energy
properties of the Kondo problem with the number of channels exceeding 2 is
governed by an intermediate-coupling fixed point, which gives rise to a
specific non-Fermi-liquid behavior of the susceptibility and specific
heat.\cite{NB,bethe,Affleck,Fabrizio} In Appendix~\ref{4-channel} we show
how one can study the properties of the intermediate-coupling fixed point
starting from the vicinity of the strong-coupling point. We discover that
on resonance the thermodynamics of our problem is indeed identical to that
of the 4-channel Kondo problem.

The technique presented in Appendix~\ref{4-channel}, however, does not
enable us to find the low-temperature conductance $G$. Nevertheless one
can estimate the order of magnitude of $G$ at the intermediate-coupling
fixed point in the following way. From the fact that the strong-coupling
fixed point is unstable and from the result (\ref{2nd_order_with_spins})
of the perturbation theory near this point we know that the conductance
must be smaller than $e^2/2\pi\hbar$. On the other hand, at $G\ll
e^2/\hbar$ the weak-coupling treatment of Sec.~\ref{weak} is applicable
and predicts the growth of conductance at $T\to0$. Therefore the limiting
value of conductance must be of the order, but smaller than
$e^2/2\pi\hbar$. We further conjecture that the peak conductance is a
universal characteristic of the intermediate-coupling fixed point, see
Appendix~\ref{4-channel}, meaning that in the limit $T\to0$ the peak
conductance does not depend on the heights of the barriers.

The scaling analysis in the spirit of Ref.~\onlinecite{Kane} enables us to
make some conclusions about the shape of the peaks in $G(N)$. In the
spinless case we see from Eq.~(\ref{conductance_spinless}) that at the
tails of the peak at $N=\frac12$ the conductance decays as $G\propto
1/\bigl(N-\frac12\bigr)^4$.  This result can be interpreted within the
scaling picture as follows. A small deviation of $N$ from $\frac12$ is
identified as a magnetic field $h$ in the appropriate multichannel Kondo
problem, see Eq.~(\ref{Kondo}).  Therefore such a deviation is a relevant
perturbation with scaling dimension $d=2/(2+k)$, where $k$ is the number
of channels.\cite{Affleck} In the spinless case we have $k=2$ and
$d=\frac12$. This means that as the temperature is lowered, the magnetic
field grows as $h(T)\propto \bigl(N-\frac12\bigr)T^{-1/2}$. We now
conjecture that near the fixed point, at $T\to0$, the conductance is a
universal function of $h$ only, $G=G(h(T))$.  As we discussed above, away
from the center of the peak $G\propto T^2$ at $T\to0$. Therefore the
magnetic field dependence of the conductance must be given by $G\propto
1/h^4\propto T^2/\bigl(N-\frac12\bigr)^4$, in agreement with
Eq.~(\ref{conductance_spinless}). We can now apply the same argument to
the case of electrons with spins, where the number of channels $k=4$ and
the scaling dimension of the magnetic field is $d=\frac13$. The
temperature dependence of the renormalized magnetic field is consequently
$h(T) \propto \bigl(N-\frac12\bigr)T^{-2/3}$.  Therefore near the
intermediate-coupling fixed point the condition $G\propto T^2$ at $T\to 0$
demands $G\propto h^{-3}\propto T^2/\bigl(N-\frac12\bigr)^3$. Thus we
expect the tails of the conductance peaks to decay as
$1/\bigl(N-\frac12\bigr)^3$.

So far we discussed the case of symmetric barriers. The on-resonance
behavior of the system is very different if the two barriers are not
identical. This can be readily seem from the analogy to the Kondo problem.
{}From the theory of the multichannel Kondo model\cite{NB,Melnikov} it is
known that the asymmetry between the channels is a relevant perturbation.
In terms of the Coulomb blockade problem this means that on resonance a
small asymmetry will grow and eventually the smaller transmission
coefficient, say ${\cal T}_L$, will scale to zero, whereas the larger one,
${\cal T}_R$, must approach the fixed point of the problem with half the
number of channels of the original one. Thus for both 2- and 4-channel
cases ${\cal T}_R \to 1$.  For the most realistic case of electrons with
spins, near this asymmetric fixed point the system is identical to the one
considered in Sec.~\ref{asymmetric}. Thus we conclude that for asymmetric
systems (i) the on-resonance conductance vanishes linearly in $T$, and (ii)
the shape of the resonances in $G(N)$ is universal and given by
Eq.~(\ref{conductance_asymmetric_final}).

\begin{figure}
\narrowtext
\epsfxsize=2.6in
\hspace{.1em}
\epsffile{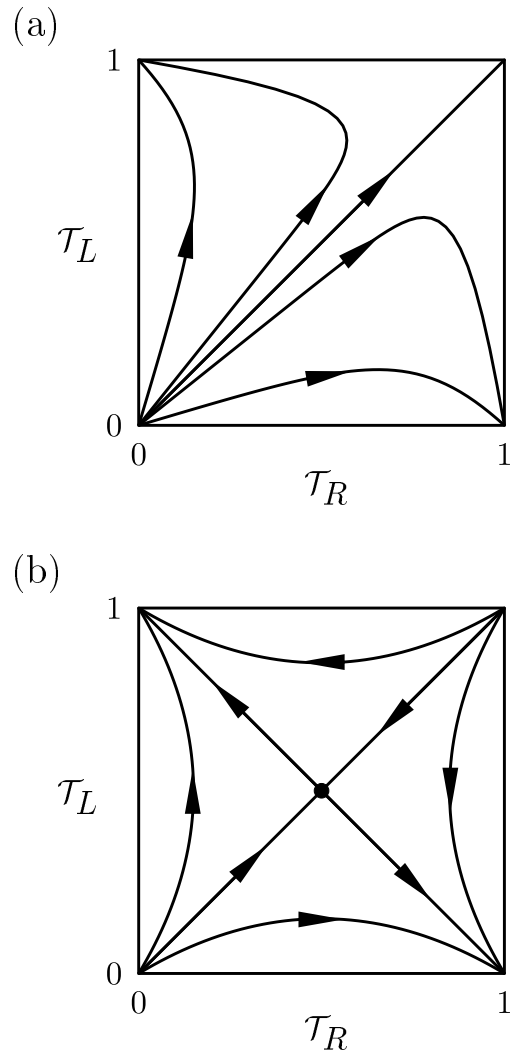}
\vspace{1.5\baselineskip}
\caption{Scaling diagram of the two-barrier system in the cases of (a)
  spinless electrons and (b) electrons with spins. The gate voltage
  corresponds to a resonance value, $N=\pm\frac12,\pm\frac32,\ldots$. In
  the symmetric case, ${\cal T}_L={\cal T}_R$, the system scales to either
  strong- or intermediate-coupling fixed point, depending on the number of
  channels. In an asymmetric case the system approaches the limit where
  the weaker barrier disappears, while the stronger one becomes very
  high.}
\label{fig:4}
\end{figure}

The on-resonance scaling properties of our problem are summarized in
Fig.~\ref{fig:4}.

\vfill

\section{Conclusion}

In this paper we have studied the transport through a quantum dot
connected to two leads by quantum point contacts, Fig.~\ref{fig:1}. Unlike
in conventional theories of Coulomb blockade,\cite{Averin,Grabert} we
concentrate on the regime of strong tunneling, where one or both contacts
are close to perfect transmission.  We find the linear conductance in a
wide interval of temperatures, limited by the level spacing in the dot
from below and by the charging energy $E_C$ from above.

We have shown that as long as there is any backscattering in the contacts,
i.e., both transmission coefficients ${\cal T}_{L,R}<1$, the Coulomb
blockade peaks in the conductance as a function of the gate voltage should
be observed. Between the Coulomb blockade peaks the temperature dependence
of the conductance at $T\to0$ is determined by the weak-coupling fixed
point. As a result we get the $T^2$ dependence of the conductance, which
coincides with the known result\cite{AN} for the inelastic co-tunneling
mechanism.

Our results show that the behavior of the conductance near the peaks
depends qualitatively on the presence of electron spins. In the spinless
case the strong-tunneling fixed point is stable if the barriers are
symmetric, and thus the technique developed in this paper allows a
complete solution of the problem. The resulting peaks in conductance have
the height $e^2/4\pi\hbar$, with the shape given by
Eq.~(\ref{conductance_spinless}). In the presence of the spins, the
strong-tunneling fixed point is no longer stable, and our technique does
not allow a complete solution of the problem. Using a useful analogy with
the multichannel Kondo problem, we have concluded that the peak
conductance must still be of the order of $e^2/\hbar$. In addition, we
studied the peaks in conductance in the case of asymmetric
barriers. The peak conductance in this case is linear in temperature at
$T\to0$, and its shape is given by
Eq.~(\ref{conductance_asymmetric_final}).

The strong-tunneling regime of Coulomb blockade was explored to some
extent in several recent experiments.\cite{van,Pasquier,Waugh,Molenkamp}
Our results are in qualitative agreement with the experiment.\cite{van} We
believe that our predictions can be easily tested in similar experiments.

\acknowledgements

We are grateful to L.~I.~Glazman and P.~A.~Lee for helpful discussions.
The work was sponsored by Joint Services Electronics Program Contract
DAAH04-95-1-0038.

\end{multicols}
\widetext

\appendix

\section{Perturbation theory for weak reflection}
\label{perturbation_theory_appendix}

In this appendix we derive the leading-order correction to the conductance
due to the weak reflection at the quantum point contacts.  We shall use a
path integral method to calculate the current-current correlation function
at imaginary frequencies and then perform analytic continuation to the
real frequencies.

\subsection{Spinless fermions}
\label{A1}

In the spinless case our problem is similar to the resonant tunneling
problem studied in Ref.~\onlinecite{Furusaki}.  Thus we can apply the
method developed in Refs.~\onlinecite{Furusaki} and \onlinecite{Furusaki2}
to calculate the conductance perturbatively with respect to the reflection
coefficients.

The current operator is given by
$I=(e/\sqrt{2}\pi)\partial_t\phi_I(0,t)$.
Then the Kubo formula (\ref{Kubo}) for the conductance can be rewritten as
\begin{equation}
G=\frac{e^2T}{2\pi^2i\hbar^2}\lim_{\omega\to 0}\omega
\lim_{i\omega_n\to\omega+i0^+}
\langle\phi_I(-i\omega_n)\phi_I(i\omega_n)\rangle,
\label{appen:Kubo}
\end{equation}
where $\omega_n=2\pi nT/\hbar$ and
\begin{equation}
\phi_I(i\omega_n)=\int^{\hbar/T}_0e^{i\omega_n\tau}\phi_I(0,\tau)d\tau.
\end{equation}
The thermal average
$\Phi_I(i\omega_n)\equiv\langle\phi_I(-i\omega_n)\phi_I(i\omega_n)\rangle$
is calculated as
\begin{equation}
\Phi_I(i\omega_n)=
\left.\frac{1}{Z^{(1)}_J}\frac{\delta^2Z^{(1)}_J}
{\delta J(i\omega_n)\delta J(-i\omega_n)}\right|_{J=0},
\end{equation}
where the generating functional is given by
\begin{eqnarray}
&&Z^{(1)}_J=
\int{\cal D}\phi_I(x,\tau)\int{\cal D}\phi_C(x,\tau)
\exp\!\left[-\frac{S_1}{\hbar}
            +\sum_{\omega_n}J(-i\omega_n)\phi_I(i\omega_n)\right],
\nonumber\\
&&S_1=
\frac{\hbar}{2\pi}\int^{\hbar/T}_0\!\!d\tau\int dx\left[
  \frac{1}{v_F}(\partial_\tau\phi_I)^2
  +v_F(\partial_x\phi_I)^2
  +\frac{1}{v_F}(\partial_\tau\phi_C)^2
  +v_F(\partial_x\phi_C)^2\right]
+\frac{2E_C}{\pi^2}\int^{\hbar/T}_0\!\!d\tau
  \left[\phi_C(0,\tau)-\frac{\pi N}{\sqrt{2}}\right]^2
\nonumber\\
&&\hspace*{8mm}
+\frac{D}{\pi}\!\int^{\hbar/T}_0\!\!d\tau\!\left(
  |r_L|\cos\bigl\{\sqrt{2}\,[\phi_I(0,\tau)-\phi_C(0,\tau)]\bigr\}\!
  +|r_R|\cos\bigl\{\sqrt{2}\,[\phi_I(0,\tau)+\phi_C(0,\tau)]\bigr\}
 \right).
\end{eqnarray}
Expanding $\exp(-S_1/\hbar)$ in powers of $|r|$ and performing Gaussian
integrals, we get
\begin{eqnarray}
\Phi_I(i\omega_n)&=&
\frac{\pi\hbar}{2|\omega_n|T}
-\frac{\gamma E_C}{2\hbar\omega^2_n}
 \bigl[|r_L|^2+|r_R|^2+2|r_L||r_R|\cos(2\pi N)\bigr]
 \int^{\hbar/T}_0\frac{1-\cos\omega_n\tau}{\sin(\pi T\tau/\hbar)}d\tau
\nonumber\\
&&
-\frac{\pi^4\gamma T^2}{8\hbar\omega^2_nE_C}
 \bigl[|r_L|^2+|r_R|^2-2|r_L||r_R|\cos(2\pi N)\bigr]
 \int^{\hbar/T}_0\frac{1-\cos\omega_n\tau}{\sin^3(\pi T\tau/\hbar)}d\tau
+{\cal O}(|r|^4).
\label{phi_Iphi_I}
\end{eqnarray}
In this calculation we have used the following relations:
\begin{mathletters}
\begin{eqnarray}
&&
\frac{T}{\hbar}\sum_{\omega'_n}
\frac{e^{-\hbar|\omega'_n|/D}}{|\omega'_n|}
(1-\cos\omega'_n\tau)
=\frac{1}{\pi}
 \ln\!\left[\frac{D}{\pi T}
            \sin\!\left(\frac{\pi T\tau}{\hbar}\right)\right],
\label{sum1}\\
&&
\frac{T}{\hbar}\sum_{\omega'_n}
\frac{e^{-\hbar|\omega'_n|/D}}{|\omega'_n|+(2E_C/\pi\hbar)}
=\frac{1}{\pi}\ln\!\left(\frac{\pi D}{2\gamma E_C}\right),
\label{sum2}\\
&&
\frac{T}{\hbar}\sum_{\omega'_n}
\frac{e^{-\hbar|\omega'_n|/D}}{|\omega'_n|+(2E_C/\pi\hbar)}
\cos\omega'_n\tau
=\frac{\pi^3T^2}{4E^2_C}\frac{1}{\sin^2(\pi T\tau/\hbar)},
\label{sum3}
\end{eqnarray}
\end{mathletters}\noindent
where we have neglected higher-order terms in $T/E_C$.
Note that Eqs.~(\ref{sum1}) and (\ref{sum3}) are correct
only for $\tau,\hbar T^{-1}-\tau\gg\hbar/D$.

In calculating the integrals over $\tau$ in Eq.~(\ref{phi_Iphi_I}), we
introduce a short-time cutoff $\tau_c\sim\hbar/D$, which is omitted in
Eqs.~(\ref{sum1}) and (\ref{sum3}), and modify the contour of integration:
\begin{equation}
\int^{\hbar/T}_0\!\!
\frac{1-\cos\omega_n\tau}{\sin^\nu(\pi T\tau/\hbar)}
d\tau=
\int^{\hbar/T-\tau_c}_{\tau_c}
\frac{1-e^{i\omega_n\tau}}{\sin^\nu(\pi T\tau/\hbar)}
d\tau=
i\!\int^\infty_0\!
 \frac{1-e^{i\omega_n(\tau_c+it)}}
      {\sin^\nu[\pi T(\tau_c+it)/\hbar]}dt
-i\!\int^\infty_0\!
 \frac{1-e^{i\omega_n(-\tau_c+it)}}
      {\sin^\nu[\pi T(\tau_c-it)/\hbar]}dt.
\label{contour}
\end{equation}
By replacing $i\omega_n$ by $\omega$ and taking the limit $\omega\to0$,
Eq.~(\ref{contour}) is reduced to $-i\omega F(T,\nu)$, where
\begin{equation}
F(T,\nu)\equiv
\int^\infty_{-\infty}
  \frac{\tau_c+it}{\sin^\nu[\pi T(\tau_c+it)/\hbar]}dt
=\frac{\hbar^2}{2\pi T^2}
 \int^\infty_{-\infty}\frac{dx}{\cosh^\nu x}
=\frac{\hbar^2}{2\pi T^2}
  \frac{\sqrt{\pi}\,\Gamma(\frac{\nu}{2})}
       {\Gamma(\frac{\nu}{2}+\frac{1}{2})}.
\label{analytic}
\end{equation}
{}From Eqs.~(\ref{appen:Kubo}), (\ref{phi_Iphi_I}), and
(\ref{analytic}) we finally obtain
\begin{equation}
G=\frac{e^2}{4\pi\hbar}\!\left\{
1-\frac{\gamma E_C}{2\pi T}
   \bigl[|r_L|^2+|r_R|^2+2|r_L||r_R|\cos(2\pi N)\bigr]
-\frac{\pi^3\gamma T}{16E_C}
   \bigl[|r_L|^2+|r_R|^2-2|r_L||r_R|\cos(2\pi N)\bigr]\right\}\!.
\label{finally_obtain}
\end{equation}
The term proportional to $T/E_C$ originates from the weak fluctuations of
the charge of the dot at low frequencies. At a half-integer $N$ it is
related to the leading irrelevant operator near the fixed point of the
2-channel Kondo problem. The scaling dimension of this operator is
$d=\frac32$ (see Sec.~\ref{scaling}), yielding the temperature dependence
$\delta G \propto T^{2d - 2} = T$.

\subsection{Electrons with spins}
\label{A2}

The calculation proceeds in parallel to the spinless case.
We first define the new sets of phase fields,
\begin{equation}
\phi_{Lc}=\frac{1}{\sqrt{2}}(\phi_{L\uparrow}+\phi_{L\downarrow}),
\quad
\phi_{Ls}=\frac{1}{\sqrt{2}}(\phi_{L\uparrow}-\phi_{L\downarrow}),
\quad
\phi_{Rc}=\frac{1}{\sqrt{2}}(\phi_{R\uparrow}+\phi_{R\downarrow}),
\quad
\phi_{Rs}=\frac{1}{\sqrt{2}}(\phi_{R\uparrow}-\phi_{R\downarrow}),
\label{phi_Lc}
\end{equation}
where $\phi_{Lc}$ and $\phi_{Rc}$ describe the charge density
fluctuations, and $\phi_{Ls}$ and $\phi_{Rs}$ correspond to the spin
density fluctuations.  We further define their momentum conjugates
$\Pi_{Lc}$, $\Pi_{Ls}$, $\Pi_{Rc}$, and $\Pi_{Rs}$ in the same way.

The current operator in this basis is $I=(e/\sqrt{2}\,\pi)
\partial_t[\phi_{Lc}(0,t)+\phi_{Rc}(0,t)]$,
and accordingly the Kubo formula for the conductance is
\begin{equation}
G=\frac{e^2T}{2\pi^2i\hbar^2}\lim_{\omega\to0}\omega
  \lim_{i\omega_n\to\omega+i0^+}\Phi_c(i\omega_n),
\label{appen:Kubo2}
\end{equation}
where $\Phi_c(i\omega_n)$ is given by
\begin{equation}
\left.
\Phi_c(i\omega_n)=
\frac{1}{Z^{(2)}_J}
\frac{\delta^2Z^{(2)}_J}{\delta J(-i\omega_n)\delta J(i\omega_n)}
\right|_{J=0}
\end{equation}
with the generating functional,
\begin{eqnarray}
&&Z^{(2)}_J=
\int\!\!{\cal D}\phi_{Lc}(x,\tau)\int\!\!{\cal D}\phi_{Ls}(x,\tau)
\int\!\!{\cal D}\phi_{Rc}(x,\tau)\int\!\!{\cal D}\phi_{Rs}(x,\tau)
\exp\!\left[-\frac{1}{\hbar}(S_2+S_J)\right],\label{Z_J}\\
&&S_2=
\frac{\hbar}{2\pi}\int^{\hbar/T}_0\!\!d\tau\int\!dx\left[
  \frac{1}{v_F}(\partial_\tau\phi_{Lc})^2+v_F(\partial_x\phi_{Lc})^2
  +\frac{1}{v_F}(\partial_\tau\phi_{Ls})^2+v_F(\partial_x\phi_{Ls})^2
\right.\nonumber\\
&&\hspace*{3.6cm}\left.
  +\frac{1}{v_F}(\partial_\tau\phi_{Rc})^2+v_F(\partial_x\phi_{Rc})^2
  +\frac{1}{v_F}(\partial_\tau\phi_{Rs})^2+v_F(\partial_x\phi_{Rs})^2
\right]\nonumber\\
&&\hspace*{8mm}
+\frac{2E_C}{\pi^2}\int^{\hbar/T}_0\!\!d\tau\left[
  \phi_{Rc}(0,\tau)-\phi_{Lc}(0,\tau)-\frac{\pi N}{\sqrt{2}}
  \right]^2\nonumber\\
&&\hspace*{8mm}
+\frac{2D}{\pi}\int^{\hbar/T}_0\!\!d\tau\left\{
  |r_L|\cos\bigl[\sqrt{2}\phi_{Lc}(0,\tau)\bigr]
       \cos\bigl[\sqrt{2}\phi_{Ls}(0,\tau)\bigr]
 +|r_R|\cos\bigl[\sqrt{2}\phi_{Rc}(0,\tau)\bigr]
       \cos\bigl[\sqrt{2}\phi_{Rs}(0,\tau)\bigr]
\right\},
\\
&&S_J=
\hbar\sum_{\omega_n}J(-i\omega_n)
  [\phi_{Lc}(i\omega_n)+\phi_{Rc}(i\omega_n)].
\nonumber
\end{eqnarray}
Up to the order $|r|^2$ the correlation function $\Phi_c(i\omega_n)$
is obtained as
\begin{eqnarray}
\Phi_c(i\omega_n)&=&
\frac{\pi\hbar}{|\omega_n|T}
-\frac{\pi}{\hbar\omega^2_n}\sqrt{\gamma TE_C}
 (|r_L|^2+|r_R|^2)\int^{\hbar/T}_0\!
   \frac{1-\cos\omega_n\tau}
        {\sin^{3/2}(\pi T\tau/\hbar)}d\tau,
\label{Phi_c}
\end{eqnarray}
where higher-order terms in $T/E_C$ are neglected.
Note that there is no term proportional to $|r_L||r_R|$ in
Eq.~(\ref{Phi_c}) because
$\langle\cos\phi_{Ls}(0)\rangle=\langle\cos\phi_{Rs}(0)\rangle=0$.
This is an important difference between the two-channel case
and four-channel case, implying that the strong-tunneling limit
is unstable even on resonance in the latter case.
{}From Eqs.~(\ref{contour}), (\ref{analytic}), (\ref{appen:Kubo2}), and
(\ref{Phi_c}) we can readily obtain Eq.~(\ref{2nd_order_with_spins}).

\subsection{Asymmetric limit}
\label{A3}

We have discussed the case of asymmetric barriers in
Sec.~\ref{asymmetric}, where the conductance is calculated in lowest order
in $v_t^{}$ but in all orders in $|r_R|$ by effectively summing up all the
most divergent terms in perturbation series.  Here we shall calculate the
leading term in the series, which diverges at low temperatures.

We will calculate the conductance in a way slightly different from the one
used in Sec.~\ref{asymmetric}.  We first note that since the left barrier
is very high, the left phase fields are pinned at the minima of the potential
$\cos\bigl(\sqrt{2}\phi_{Lc}\bigr)\cos\bigl(\sqrt{2}\phi_{Ls}\bigr)$, i.e.,
$(\phi_{Lc},\phi_{Ls})=\biglb(\sqrt{2}\pi(m+\frac{1}{2}),\sqrt{2}\pi
n\bigrb)$ or $\biglb(\sqrt{2}\pi m,\sqrt{2}\pi(n+\frac{1}{2})\bigrb)$,
where $m$ and $n$ are integers.\cite{Furusaki,Furusaki2} Thus the tunneling
of an electron through the left barrier corresponds to a sudden change in
the phase fields, say, $(\phi_{Lc},\phi_{Ls})=(\pi/\sqrt2,0)\to(\sqrt2\pi,
\pi/\sqrt2)$.  The probability for this tunneling process is then given by
\begin{equation}
P_+=\frac{|\tilde v_t^{}|^2}{\hbar^2}\int^\infty_{-\infty}
\bigl\langle e^{iH_it/\hbar}e^{-iH_ft/\hbar}\bigr\rangle_i e^{ieVt/\hbar}dt,
\label{P_+}
\end{equation}
where $\tilde v_t^{}$ is a constant proportional to $v_t$, $V$ is a
voltage across the left barrier, and $H_i$ and $H_f$ are the bosonized
Hamiltonian (\ref{H_0_with_spins})--(\ref{backscattering_with_spins}) with
the condition $\bigl(\phi_{Lc}(0),\phi_{Ls}(0)\bigr)=(\pi/\sqrt2,0)$ and
$(\sqrt2\pi,\pi/\sqrt2)$, respectively.  The thermal average
$\langle\ldots\rangle_i$ is taken for the Hamiltonian $H_i$.  It is clear
that the probability for the inverse tunneling process is given by
$P_-=e^{-eV/T}P_+$. Since the current is $I=2e(P_+-P_-)$ with the factor
$2$ coming from electron spins, the linear conductance is given by
\begin{equation}
G=\frac{2e^2|\tilde v_t^{}|^2}{\hbar^2T}\int^\infty_{-\infty}
\bigl\langle e^{iH_it/\hbar}e^{-iH_ft/\hbar}\bigr\rangle_i dt
=\frac{2e^2|\tilde v_t^{}|^2}{\hbar^2T}\int^\infty_{-\infty}K_0(t)dt.
\label{cond-asym}
\end{equation}
It will soon become clear that the correlator $K_0(t)$ is related to $K(t)$
in Eq.~(\ref{conductance_asymmetric}) as
$K_0(t)=\{\pi T/iD\sinh[\pi T(t-i\delta)/\hbar]\}^2K(t)$.
For the imaginary time $\tau=it$ we can calculate the correlator using
the path integral.  To get the conductance we then need to perform analytic
continuation, $\tau\to it+\delta$, and calculate the integral
(\ref{cond-asym}).

The partition function of the system
$Z^{(2)}_J|_{J=0}$ is given by Eq~(\ref{Z_J}).
Since both the charging energy and the cosine potential depend on the
phase fields at the point $x=0$ only, we can integrate out\cite{Furusaki2}
the phase fields off the point to get the following effective action:
\begin{eqnarray}
S_{\rm eff}&=&
\frac{T}{\pi}\sum_{\omega_n}|\omega_n|\bigl[
|\phi_{Lc}(i\omega_n)|^2+|\phi_{Ls}(i\omega_n)|^2
 +|\phi_{Rc}(i\omega_n)|^2+|\phi_{Rs}(i\omega_n)|^2
\bigr]\nonumber\\
&&
+\frac{2E_C}{\pi^2}\int^{\hbar/T}_0\!\!d\tau
 \left[\phi_{Rc}(\tau)-\phi_{Lc}(\tau)
       -\frac{\pi N}{\sqrt{2}}\right]^2\nonumber\\
&&
+\frac{2D}{\pi}\int^{\hbar/T}_0\!\!d\tau
 \left\{|r_L|\cos\bigl[\sqrt{2}\phi_{Lc}(\tau)\bigr]
             \cos\bigl[\sqrt{2}\phi_{Ls}(\tau)\bigr]
       +|r_R|\cos\bigl[\sqrt{2}\phi_{Rc}(\tau)\bigr]
             \cos\bigl[\sqrt{2}\phi_{Rs}(\tau)\bigr]\right\}.
\label{S_eff}
\end{eqnarray}
With this effective action we can readily calculate $K_0(-i\tau)$ as
\begin{equation}
K_0(-i\tau)=
\frac{\int{\cal D}\phi_{Rc}\int{\cal D}\phi_{Rs}
       \exp\bigl\{-\frac{1}{\hbar}S_{\rm eff}\bigl[
          \phi^{(1)}_{Lc},\phi^{(1)}_{Ls},
          \phi_{Rc},\phi_{Rs}\bigr]\bigr\}}
     {\int{\cal D}\phi_{Rc}\int{\cal D}\phi_{Rs}
       \exp\bigl\{-\frac{1}{\hbar}S_{\rm eff}\bigl[
          \phi^{(0)}_{Lc},\phi^{(0)}_{Ls},
          \phi_{Rc},\phi_{Rs}\bigr]\bigr\}},
\label{K_0(-itau)}
\end{equation}
where
\begin{mathletters}
\begin{eqnarray}
\biglb(\phi^{(0)}_{Lc}(\tau'),\phi^{(0)}_{Ls}(\tau')\bigrb)&=&
\bigl(\pi/\sqrt{2},0\bigr)\quad 0<\tau'<\hbar/T,\\
\biglb(\phi^{(1)}_{Lc}(\tau'),\phi^{(1)}_{Ls}(\tau')\bigrb)&=&
\cases{
\bigl(\sqrt{2}\pi,\pi/\sqrt{2}\,\bigr)&$0<\tau'<\tau$,\cr
\bigl(\pi/\sqrt{2},0\bigr)&$\tau<\tau'<\hbar/T$.}
\end{eqnarray}
\end{mathletters}

We shall first consider the non-interacting case, $E_C=0$, to find the
coefficient in Eq.~(\ref{cond-asym}).  In this case we may ignore the
fields $\phi_{Rc}$ and $\phi_{Rs}$.  Thus the correlator is given by
\begin{equation}
K_0(-i\tau)=
\frac{\exp\bigl\{-\frac{1}{\hbar}S_{\rm eff}\bigl[
                \phi^{(1)}_{Lc},\phi^{(1)}_{Ls},0,0\bigr]\bigr\}}
     {\exp\bigl\{-\frac{1}{\hbar}S_{\rm eff}\bigl[
                \phi^{(2)}_{Lc},\phi^{(2)}_{Ls},0,0\bigr]\bigr\}}
=\left[\frac{\pi T}{D\sin(\pi T\tau/\hbar)}\right]^2,
\label{K_0-nonint}
\end{equation}
where we have used Eq.~(\ref{sum1}).  Since we know that in this case
the conductance is $G=G_L$, we rewrite Eq.~(\ref{cond-asym}) as
\begin{equation}
G=\frac{G_LD^2}{2\pi\hbar T}\int^\infty_{-\infty}K_0(t)dt.
\label{cond-asym-2}
\end{equation}

With nonzero charging energy, $K_0(-i\tau)$ is calculated up to the order
$|r_R|^2$ as
\begin{equation}
K_0(-i\tau)=
\frac{\pi^4T^3}{\gamma E_CD^2}
\frac{1}{\sin^3(\pi T\tau/\hbar)}
\left\{
1-\frac{\Gamma_RT}{\hbar^2}
   \int^\tau_0d\tau_1\int^{\hbar/T}_\tau\!d\tau_2
   \frac{1}{\sin[\pi T(\tau_2-\tau_1)/\hbar]}
\right\},
\label{K_0(-itau)-2}
\end{equation}
where $\Gamma_R$ is defined in Eq.~(\ref{Gamma_R}).
In this calculation we have neglected terms which are smaller
in $T/E_C$ such as
$T\sum_{\omega_n}\sin(\omega_n\tau_j)[|\omega_n|+(2E_C/\pi\hbar)]^{-1}$.
For $0\le\tau\le\frac{\hbar}{2T}$ the double integral $I_2(\tau)$ is
calculated as
\begin{equation}
I_2(\tau)=
2\int^\tau_0\frac{\tau_0}{\sin(\pi T\tau_0/\hbar)}d\tau_0
-2\tau\int^{\hbar/2T}_\tau\frac{d\tau_0}{\sin(\pi T\tau_0/\hbar)}
=
\frac{i\hbar}{T}\left(\frac{\hbar}{2T}-\tau\right)
+2\int^\infty_0\frac{t_0}{\sin[\pi T(\tau+it_0)/\hbar]}dt_0.
\end{equation}
To get the conductance we need the following integral:
\begin{equation}
\int^\infty_{-\infty}\frac{I_2(it)}{\sin^3(i\pi Tt/\hbar)}dt
=
2\left(\frac{\hbar}{\pi T}\right)^3
\int^\infty_{-\infty}\!dx\int^\infty_0\!dy
\frac{y}{\cosh^3x\cosh(x+y)}
=
\left(\frac{\hbar}{\pi T}\right)^3[7\zeta(3)-2].
\label{int-zeta}
\end{equation}
{}From Eqs.~(\ref{cond-asym-2}), (\ref{K_0(-itau)-2}), and (\ref{int-zeta})
we get
\begin{equation}
G=G_L\frac{\pi^3T}{8\gamma E_C}
\left\{1-\frac{2\Gamma_R}{\pi^3T}[7\zeta(3)-2]\right\}.
\label{zeta}
\end{equation}
The leading correction due to the weak reflection in the right contact
diverges\cite{Flensberg} at low temperatures unless $N$ is half-integer.
We note that for $T\gg\Gamma_R$ Eq.~(\ref{conductance_asymmetric_final})
in fact reduces to Eq.~(\ref{zeta}).

The above result is for the case of electrons with spins.  We can, of
course, perform the same kind of calculation for the spinless case as
well.  Here we give only the final result:
\begin{equation}
G = G_L \frac{2\pi^4 T^2} {3\gamma^2E_C^2} [1-4\gamma|r_R|\cos(2\pi N)].
\label{G-asym-spinless}
\end{equation}
There are two important differences between Eqs.~(\ref{zeta}) and
(\ref{G-asym-spinless}). First, even without the reflection in the right
contact, the conductance (\ref{G-asym-spinless}) is already proportional
to $T^2$. Second, the leading correction gives no additional temperature
dependence.  Thus the perturbation expansion in powers of $|r_R|$ is well
behaved for the spinless case, in striking contrast with the case with
spins.  A similar difference between the two cases was found for the
charge of a quantum dot with a single contact.\cite{Matveev1}

\section{4-channel Kondo fixed point approached from the strong
         coupling limit}
\label{4-channel}

In this appendix we demonstrate that, when $N$ is half-integer (on
resonance), our 1D model is closely related to the 4-channel Kondo model
by transforming our bosonized Hamiltonian to the one that recently appeared
in the study of 4-channel Kondo problem.\cite{Fabrizio}

First we rewrite the Hamiltonian
(\ref{H_0_with_spins})--(\ref{backscattering_with_spins}) with the fields
introduced in Eq.~(\ref{phi_Lc}).  We then introduce new
fields $\varphi_{Lc}^{}$, $\varphi_{Ls}^{}$, $\varphi_{Rc}^{}$, and
$\varphi_{Rs}^{}$ defined in the same way as $\varphi_+$ in
Eq.~(\ref{chiralization}), each describing the even modes of $\phi_{Lc}$,
$\phi_{Ls}$, $\phi_{Rc}$, and $\phi_{Rs}$.  The Hamiltonian is now given
by
\begin{eqnarray}
H_0&=&\frac{v_F}{4\pi}\int\left\{
[\nabla\varphi_{Lc}^{}(x)]^2+[\nabla\varphi_{Ls}^{}(x)]^2
+[\nabla\varphi_{Rc}^{}(x)]^2+[\nabla\varphi_{Rs}^{}(x)]^2
\right\}dx,
\label{H_0-1}
\\
H_C&=&
\frac{E_C}{\pi^2}\left[\varphi_{Rc}^{}(0)-\varphi_{Lc}^{}(0)-\pi N\right]^2,
\label{H_C-1}
\\
H_V&=&
\frac{2D|r_L|}{\pi}\cos[\varphi_{Lc}^{}(0)]\cos[\varphi_{Ls}^{}(0)]
+\frac{2D|r_R|}{\pi}\cos[\varphi_{Rc}^{}(0)]\cos[\varphi_{Rs}^{}(0)].
\label{H_V-1}
\end{eqnarray}
With these new fields the current operator is written as
\begin{equation}
I=
-\frac{ev_F}{2\pi}\nabla[\varphi_{Lc}^{}(x)+\varphi_{Rc}^{}(x)]\biggr|_{x=0}.
\label{I-1}
\end{equation}
Next we fermionize the bosonic fields as in (\ref{fermionization}) to get
\begin{eqnarray}
&&H_0=
i\hbar v_F\int\!\left[
\psi^\dagger_{Lc}(x)\nabla\psi_{Lc}^{}(x)
+\psi^\dagger_{Ls}(x)\nabla\psi_{Ls}^{}(x)
+\psi^\dagger_{Rc}(x)\nabla\psi_{Rc}^{}(x)
+\psi^\dagger_{Rs}(x)\nabla\psi_{Rs}^{}(x)
\right]dx,
\label{H_0-2}
\\
&&H_C=
E_C\left\{\int{\rm sgn}(x)
 \left[
   \psi^\dagger_{Lc}(x)\psi_{Lc}^{}(x)-\psi^\dagger_{Rc}(x)\psi_{Rc}^{}(x)
 \right]dx
-N'\right\}^2,
\label{H_C-2}
\\
&&H_V=
-\hbar v_F|r_L|\eta_{Lc}^{}\eta_{Ls}^{}
 \left[\psi_{Lc}^{}(0)-\psi^\dagger_{Lc}(0)\right]
 \left[\psi_{Ls}^{}(0)-\psi^\dagger_{Ls}(0)\right]\nonumber\\
&&\hspace*{10mm}
-\hbar v_F|r_R|\eta_{Rc}^{}\eta_{Rs}^{}
 \left[\psi_{Rc}^{}(0)-\psi^\dagger_{Rc}(0)\right]
 \left[\psi_{Rs}^{}(0)-\psi^\dagger_{Rs}(0)\right],
\label{H_V-2}
\\
&&I=
-ev_F:\psi^\dagger_{Lc}(0)\psi_{Lc}^{}(0)
      +\psi^\dagger_{Rc}(0)\psi_{Rc}^{}(0):\,,
\label{I-2}
\end{eqnarray}
where the $\eta$'s are Majorana fermions, the $\psi$'s chiral fermions, and
$N'=N+{\rm const}$.
Since the products of the Majorana fermions, $\eta_{Lc}^{}\eta_{Ls}^{}$ and
$\eta_{Rc}^{}\eta_{Rs}^{}$, commute with each other, and
$(\eta_{Lc}^{}\eta_{Ls}^{})^2 = (\eta_{Rc}^{}\eta_{Rs}^{})^2=-1$, we can
replace them by $-i$.
We then introduce another set of fermions:
\begin{mathletters}
\begin{eqnarray}
&&\psi_{1c}^{}(x)=
\frac{1}{\sqrt{2}}\left[\psi_{Lc}^{}(x)+i\psi^\dagger_{Rc}(x)\right],
\label{psi_1c}
\\
&&\psi_{2c}^{}(x)=
\frac{1}{\sqrt{2}}\left[\psi^\dagger_{Lc}(x)+i\psi_{Rc}^{}(x)\right],
\label{psi_2c}
\\
&&\psi_s^{}(x)=
\frac{1}{2}\!\left[
\psi_{Ls}^{}(x)-\psi^\dagger_{Ls}(x)+i\psi_{Rs}^{}(x)-i\psi^\dagger_{Rs}(x)
\right],
\label{psi_1s}
\\
&&\tilde\psi_s^{}(x)=
\frac{1}{2}\!\left[
\psi_{Ls}^{}(x)+\psi^\dagger_{Ls}(x)+i\psi_{Rs}^{}(x)+i\psi^\dagger_{Rs}(x)
\right].
\label{psi_2s}
\end{eqnarray}
\end{mathletters}\noindent
In terms of these fermions the Hamiltonian and the current operator are
written as
\begin{eqnarray}
&&H_0=
i\hbar v_F\int\!\left[
\psi^\dagger_{1c}(x)\nabla\psi_{1c}^{}(x)
+\psi^\dagger_{2c}(x)\nabla\psi_{2c}^{}(x)
+\psi^\dagger_s(x)\nabla\psi_s^{}(x)
+\tilde\psi^\dagger_s(x)\nabla\tilde\psi_s^{}(x)
\right]dx,
\label{H_0-3}
\\
&&H_C=
E_C\left\{\int{\rm sgn}(x)
 \left[
 \psi^\dagger_{1c}(x)\psi_{1c}^{}(x)-\psi^\dagger_{2c}(x)\psi_{2c}^{}(x)
 \right]dx
-N'\right\}^2,
\label{H_C-3}
\\
&&H_V=
\frac{i}{\sqrt{2}}\hbar v_F|r_L|
\left[
 \psi_{1c}^{}(0)-\psi^\dagger_{1c}(0)-\psi_{2c}^{}(0)+\psi^\dagger_{2c}(0)
\right]
\left[\psi_s^{}(0)-\psi^\dagger_s(0)\right]\nonumber\\
&&\hspace*{10mm}
+\frac{i}{\sqrt{2}}\hbar v_F|r_R|
\left[
 \psi_{1c}^{}(0)+\psi^\dagger_{1c}(0)-\psi_{2c}^{}(0)-\psi^\dagger_{2c}(0)
\right]
\left[\psi_s^{}(0)+\psi^\dagger_s(0)\right],
\label{H_V-3}
\\
&&I=
-ev_F\left[
 \psi^\dagger_{1c}(0)\psi^\dagger_{2c}(0)
 +\psi_{2c}^{}(0)\psi_{1c}^{}(0)
\right].
\label{I-3}
\end{eqnarray}
The fermion field $\tilde\psi_s$ is decoupled from the other fields,
and thus we neglect it in the following.
We bosonize the remaining three fermions again:
\begin{mathletters}
\begin{eqnarray}
&&\psi_{1c}^{}(x)=
\left(\frac{D}{2\pi\hbar v_F}\right)^{1/2}\eta_{1c}^{}e^{i\varphi_1^{}(x)},
\label{eta_1c}
\\
&&\psi_{2c}^{}(x)=
\left(\frac{D}{2\pi\hbar v_F}\right)^{1/2}\eta_{2c}^{}e^{i\varphi_2^{}(x)},
\label{eta_2c}
\\
&&\psi_s^{}(x)=
\left(\frac{D}{2\pi\hbar v_F}\right)^{1/2}\eta_s^{}e^{i\varphi_s^{}(x)},
\label{eta_s}
\end{eqnarray}
\end{mathletters}\noindent
where the $\varphi$'s are bosonic fields and the $\eta$'s are Majorana
fermions.
Equations (\ref{H_0-3})--(\ref{I-3}) are then rewritten as
\begin{eqnarray}
&&H_0=
\frac{v_F}{4\pi}\int\!\left\{
[\nabla\varphi_1^{}(x)]^2+[\nabla\varphi_2^{}(x)]^2
+[\nabla\varphi_s^{}(x)]^2\right\}dx,
\label{H_0-4}
\\
&&H_C=\frac{E_C}{\pi^2}\left[
  \varphi_2^{}(0)-\varphi_1^{}(0)-\pi\left(N-\frac{1}{2}\right)
  \right]^2,
\label{H_C-4}
\\
&&H_V=
-\frac{\sqrt{2}}{\pi}D|r_L|
 \left[\eta_1^{}\sin\varphi_1^{}(0)-\eta_2^{}\sin\varphi_2^{}(0)\right]
 \sin\varphi_s^{}(0)
+\frac{\sqrt{2}}{\pi}D|r_R|
 \left[\eta_1^{}\cos\varphi_1^{}(0)-\eta_2^{}\cos\varphi_2^{}(0)\right]
 \cos\varphi_s^{}(0),
\label{H_V-4}
\\
&&
I=-i\frac{eD}{\hbar}\eta_1^{}\eta_2^{}
    \sin[\varphi_1^{}(0)+\varphi_2^{}(0)],
\label{I-4}
\end{eqnarray}
where $\eta_1^{}=i\eta_{1c}^{}\eta_s^{}$ and
$\eta_2^{}=i\eta_{2c}^{}\eta_s^{}$.
In Eq.~(\ref{H_C-4}) the parameter $N$ is shifted by $\frac{1}{2}$
so that the Hamiltonian (\ref{H_0-4})--(\ref{H_V-4}) yields the same
perturbation series for the partition function in powers of $|r|$
as the original Hamiltonian (\ref{H_0-1})--(\ref{H_V-1}).

Let us consider the case where $N$ is half-integer (on resonance)
and $|r_L|=|r_R|=r_0$ (symmetric barriers).
At low temperatures the difference $\varphi_2(0)-\varphi_1(0)$ is
almost fixed to be $\pi(N-\frac{1}{2})$ due to the charging energy.
In the strong-tunneling limit ($r_0\ll1$), we can thus take average of
the Hamiltonian over $\varphi_2-\varphi_1$ to get an effective Hamiltonian
for $\varphi_s$ and $\bar\varphi=(\varphi_1+\varphi_2)/\sqrt{2}$:
\begin{eqnarray}
&&
\langle H_0+H_V\rangle=
\frac{v_F}{4\pi}\int\!\left\{
[\nabla\bar\varphi(x)]^2+[\nabla\varphi_s^{}(x)]^2
\right\}dx
+\frac{2}{\pi}\left(\frac{\gamma E_C}{\pi D}\right)^{1/4}
 Dr_0(\eta_1^{}-\eta_2^{})
 \cos\!\left(\frac{\bar\varphi(0)}{\sqrt{2}}+\varphi_s^{}(0)\right),
\label{H-5}
\\
&&
I=-i\frac{eD}{\hbar}\eta_1^{}\eta_2^{}
    \sin\!\left[\sqrt{2}\,\bar\varphi(0)\right].
\label{I-5}
\end{eqnarray}
Introducing new bosonic fields,
$\varphi_a^{}=\sqrt{\frac{2}{3}}\bar\varphi-\sqrt{\frac{1}{3}}\varphi_s^{}$
and
$\varphi_b^{}=\sqrt{\frac{1}{3}}\bar\varphi+\sqrt{\frac{2}{3}}\varphi_s^{}$,
we rewrite the Hamiltonian and the current as
\begin{eqnarray}
&&\langle H_0+H_V\rangle=
\frac{v_F}{4\pi}\int\!\left\{
[\nabla\varphi_a^{}(x)]^2+[\nabla\varphi_b^{}(x)]^2
\right\}dx
+\frac{2\sqrt{2}}{\pi}\left(\frac{\gamma E_C}{\pi D}\right)^{1/4}
 Dr_0\sigma_x\cos\!\left[\sqrt{\frac{3}{2}}\varphi_b^{}(0)\right],
\label{H-6}
\\
&&I=
\frac{eD}{\hbar}\sigma_z
\sin\!\left[\frac{2}{\sqrt{3}}\varphi_a^{}(0)
            +\sqrt{\frac{2}{3}}\varphi_b^{}(0)\right],
\label{I-6}
\end{eqnarray}
where $\sigma_x=(\eta_1^{}-\eta_2^{})/\sqrt{2}$ and
$\sigma_z=-i\eta_1^{}\eta_2^{}$ are Pauli matrices.
The Hamiltonian (\ref{H-6}) is the same as that of a single impurity
in the $g=\frac{3}{4}$ Luttinger liquid.\cite{Kane}
Since this $g=\frac{3}{4}$ problem is related to the 4-channel Kondo
problem in the Toulouse limit,\cite{Fabrizio} the above transformations
show that  in the strong-tunneling limit our bosonized Hamiltonian is
indeed equivalent to the 4-channel Kondo problem.
The infrared stable fixed point of the 4-channel Kondo problem
corresponds to the case with $r_0\to\infty$ in Eq.~(\ref{H-6}).
At this point $\sigma_x\cos[(3/2)^{1/2}\varphi_b^{}]$ takes $-1$ so that
the field $\varphi_b^{}(0)$ is pinned at either $\varphi_b^{}(0)=\pm 2n
(2/3)^{1/2}\pi$ (for $\sigma_x=-1$) or $\varphi_b^{}(0)-\pi=\pm 2m(2/3)^{1/2}
\pi$ (for $\sigma_x=1$), where $m$ and $n$ are integers.  Thus in calculating
the dimension of the current operator (\ref{I-6}) we may neglect
$\varphi_b^{}(0)$. We then immediately see that the dimension is $d=1$
because the operator $\exp[i(4/3)^{1/2}\varphi_a(0)]$ has dimension $2/3$
and the dimension of $\sigma_z$ is\cite{Fabrizio} $1/3$. This means that
the leading term of the conductance is temperature-independent at $T\to0$.
To calculate the conductance we need to know the coefficient $\lambda$ of the
correlator $\langle[I(t),I(0)]\rangle=-\lambda e^2/t^2$ at $T\to0$, which is a
difficult problem. We expect, however, that $\lambda$ should be universal,
implying that at $T=0$ the linear conductance $G$ has a universal value
$\pi \lambda e^2/2\hbar$ independent of the initial value of $r_0$.

\begin{multicols}{2}

\end{multicols}


\begin{references}
\bibitem{remark} Strictly speaking, the presence of the Majorana fermion
  $\eta_{+}^{}$ in Eq.~(\ref{fermionization}) is not necessary, since we
  have only one bosonic branch. Nevertheless, it is more convenient to add
  $\eta_{+}^{}$, so that the resulting Hamiltonian (\ref{H_fermionized})
  had only terms quadratic in fermion operators.
\bibitem{remark2} A formal proof can be obtained in a way similar to the
  derivation (\ref{unitary_transformation})--(\ref{transformed_H_C}) of
  the expression $K(t)=K_\Theta(t)$ for the correlator (\ref{K}). Instead
  of the transformation (\ref{unitary_transformation}) one should use
  $U=(-1)^{\hat n c^\dagger c}$.
\end{references}
\end{document}